\documentclass[pdftex,12pt]{article}
\usepackage{jheppub,amsmath,amsfonts,amsbsy,latexsym,jheppub,amssymb,amscd,amstext}
\usepackage{pst-node}
\usepackage{dsfont}
\usepackage{overpic,youngtab}
\pdfoutput=1
\usepackage{rotating}
\usepackage{tikz}
\usepackage{slashed}
\pdfoutput=1

\def\bpm{\begin{pmatrix}}
	\def\epm{\end{pmatrix}}
\def\be{\begin{equation}}
\def\ee{\end{equation}}
\def\bea{\begin{eqnarray}}
\def\eea{\end{eqnarray}}
\def\pd{\partial}
\def\a{\alpha}

\def\b{\beta}

\def\g{\gamma}
\def\d{\delta}
\def\m{\mu}

\def\n{\nu}
\def\t{\tau}

\def\l{\lambda}

\def\k{\kappa}
\def\r{\rho}

\def\bR{\bar{R}}

\def\zp{z^\prime}
\def\yp{y^\prime}
\def\bg{\bar{g}}

\def\xp{x^\prime}

\def\bp{\bar{\phi}}
\def\s{\sigma}
\def\e{\epsilon}

\def\bma{\begin{pmatrix}}
	\def\ema{\end{pmatrix}}

\def\bp{\bar{\phi}}

\def\bg{\bar{g}}
\def\bi{\begin{itemize}}
	\def\ei{\end{itemize}}
\def\bn{\bar{\nabla}}

\def\bp{\bar{\phi}}
\def\tr{{\rm tr\,}}

\begin{document}

	\vspace*{-1cm}
	\phantom{hep-ph/***} 
	{\flushleft
		{{FTUAM-20-23}}
		\hfill{{ IFT-UAM/CSIC-20-151}}}
	\vskip 1.5cm
	\begin{center}
		{\LARGE\bfseries Weighing the Vacuum  Energy}\\[3mm]
		\vskip .3cm
		
	\end{center}

	\vskip 0.5  cm
	\begin{center}
		{\large Enrique Alvarez, Jesus Anero and Raquel Santos-Garcia }
		\\
		\vskip .7cm
		{
			Departamento de F\'isica Te\'orica and Instituto de F\'{\i}sica Te\'orica, 
			IFT-UAM/CSIC,\\
			Universidad Aut\'onoma de Madrid, Cantoblanco, 28049, Madrid, Spain\\
			\vskip .1cm

			\vskip .5cm
			\begin{minipage}[l]{.9\textwidth}
				\begin{center} 
					\textit{E-mail:} 
					\tt{enrique.alvarez@uam.es},
					\tt{jesusanero@gmail.com},
					\tt{raquel.santosg@uam.es}
				\end{center}
			\end{minipage}
		}
	\end{center}
	\thispagestyle{empty}
	
	\begin{abstract}
		\noindent
		We discuss the weight of vacuum energy in various contexts. First, we compute the vacuum energy for flat spacetimes of the form $\mathbb{T}^3 \times \mathbb{R}$, where $\mathbb{T}^3$ stands for a general 3-torus. We discover a quite simple relationship between energy at radius $R$ and energy at radius  $\frac{l_s^2}{ R}$. Then we consider quantum gravity effects in the vacuum energy of a scalar field in $\mathbb{M}_3 \times S^1$ where $\mathbb{M}_3$ is a general curved spacetime, and the circle $S^1$ refers to a spacelike coordinate.  We compute it for General Relativity and generic transverse {\em TDiff} theories. In the particular case of Unimodular Gravity vacuum energy does not gravitate. 
		%find an unambiguous energy momentum tensor for the vacuum energy.
	\end{abstract}
	
	\newpage
	\tableofcontents
	\thispagestyle{empty}
	\thispagestyle{empty}
	\newpage
	\setcounter{page}{1}

	%\newpage
	%%%%%%%%%%%%%%%
	
	%\tableofcontents
	
	\newpage
	%%%%%%%%%%%%%%%%%%%%%%%%%%%%%%%%%%%%%%%%%%%%%%%%%%%%%%%%%%%%%%%%%%%%%%%%%%%%%%%%%%%%%%%%%%%%%%%%%%%%%%%%%%%%%%%
	\section{Introduction}
	%%%%%%%%%%%%%%%%%%%%%%%%%%%%%%%%%%%%%%%%%%%%%%%%%%%%%%%%%%%%%%%%%%%%%%%%%%%%%%%%%%%%%%%%%%%%%%%%%%%%%%%%%%%%%%%
	The existence of vacuum energy is a prediction of quantum field theory (QFT), although explicit computations usually yield a divergent value for this observable. This is not a problem whenever the gravitational interaction can be neglected, because then the zero-point energy is physically irrelevant and some normal ordering can be imposed which renormalizes the vacuum energy to zero.  %in which the vacuum energy is renormalized to nil. 
	This situation changes, however, once the effects of the gravitational field are taken into account. Then the vacuum energy weighs, and its renormalization is physically relevant.
	\par
	There are different senses in which we can speak about vacuum energy (cf. the seminal paper on the Casimir effect \cite{Casimir} and related comments in \cite{Blau})\footnote{ We refer to \cite{Plunien,Plunien2} for reviews as well as to \cite{Fulling:1989} for a theoretical treatment.}
	These ambiguities are not unrelated with recent concerns on how the said Casimir energy falls in an external gravitational field; that is, whether or not it violates the equivalence principle (see \cite{Fulling,Fulling2} and references therein).  The main issue follows from the use of the energy-momentum tensor to infer the vacuum energy via the following variational formula
	\be
	\d W=-\dfrac{1}{2}\int \sqrt{|g|}\,d^nx  \, T^{\m\n} \d g_{\m\n}.
	\ee
	Ambiguities arise because the computed energy-momentum tensor is not conserved. This means that the above expression is not gauge invariant.
	In fact, in almost all treatments known to us, the gravitational field is considered as a background field and the Casimir effect is encapsulated in some energy-momentum tensor (vacuum energy density). The treatment in \cite{Huggins,odintsov1,odintsov4,odintsov2,Cho} is an exception as it is an example of how to compute the gauge invariant Vilkovisky-DeWitt effective action of quantum gravity.
	%More on this later.
	\par
	It is worth pointing out the work of Jaffe and coworkers \cite{Jaffe,Jaffe2} that claim (rightly so in our opinion) that the experiments made up to now do not test the reality of the {\em vacuum energy}, but rather of the {\em Casimir force} which can be computed (as they do) using standard scattering techniques. Nevertheless, these experiments by themselves do not tell us anything about the weight (if any) of the vacuum energy. Incidentally, one of the first persons to worry about this subject, namely Pauli \cite{Pauli} , denied the physical relevance of the vacuum energy and claimed that it should be subtracted from the total energy-momentum of the system. 
	\par
	Our definition of vacuum energy stems from the background field approach in QFT. When the gravitational field is treated as a gauge field then the effective action, when all background matter fields are taken to be zero, contains a leading term of the form
	\be
	W_0 = \int \sqrt{|g|}\,d^n x \,\mathcal{E}_0,
	\ee
	where $\mathcal{E}_0$ is the {\em constant vacuum energy density}, that is, the {\em cosmological constant}; the field-independent piece of the effective potential. With this definition, the engineering dimension of $\mathcal{E}_0$ is $n$. Other  definitions are often used in the literature, and it is usually easy to relate them to our $W_0$. In particular, we shall sometimes use the notation $E_0$ for one such quantity with mass dimension one.
	\par
	It is also interesting  to consider some modifications of General Relativity, namely {\em transverse theories} in which the volume element is changed to
	\be
	d_T(vol)\equiv f(g) \, d^n x,
	\ee
	where $f(g)$ is an arbitrary function of the determinant of the metric tensor. We shall eventually comment on the particular case of Unimodular Gravity, in which this $f(g)=1$, so that the same term reads
	\be
	W = \int\,d^nx\, \mathcal{E}_0.
	\ee
	As a consequence, the vacuum energy density does not weigh through a direct coupling with the gravitational field.  A similar coupling is indeed necessary owing to self-consistency (i.e. Bianchi identity), but the point is that its effect is {\em not} proportional to the constant $\mathcal{E}_0$.
	\par
	Let us now summarize the contents of this paper.
	After reviewing the standard treatment of vacuum energy in flat space in our language,  we generalize it to more general (still background; that is, neglecting backreaction)  flat manifolds of the type $\mathbb{T}_3\times \mathbb{R}$, where the three-dimensional manifold is a general torus.
	% {\mathbb{R}^3\over \Gamma} (where $\Gamma $ is a subgroup of isometries acting freely on $\mathbb{R}^3$).  
	In this simple situation, we can unveil some relationship
	between the vacuum energy at radius $R$ and at radius $l_s^2/R$, in a sense to be clarified later. Then we proceed to study the quantum gravity effects. We assume that the background spacetime remains of the form $\mathbb{M}_3 \times S^1$, (where $\mathbb{M}_3$ is not necessarily flat) even after quantum corrections.
	%; that is, we neglect the backreaction. 
	Our treatment is gauge invariant from the very beginning, because when all interactions (including gravity) are quantized and integrated upon in the path integral, there is no other room for ambiguity than the renormalization conditions to be imposed on finite parts once appropriate counterterms are included at each order in the loop expansion.
	\par
	In that sense, as we have already pointed out, for us the vacuum energy is related to the constant term in the effective lagrangian, which in Einstein's General Relativity couples directly to gravity only through the term $\sqrt{|g|}$. This means that its effect on the energy-momentum tensor is proportional to the background spacetime metric
	\be
	T_{\m\n}^{vac}\sim f(x)g_{\m\n},
	\ee
	assuming there are no boundaries in the spacetime. This procedure circumvents the nasty task of defining {\em energy} in an arbitrary background spacetime, $\bg_{\m\n}$, although it is true that the name is only appropriate in some simple cases in which the total energy can be properly defined.
	
	\subsection{%Lightning 
		Review of known results}
%%%%%%%%%%%%%%%%%%%%%%%%%%%%%%%%%%%%%%%%%%%%%%%%%%%%%%%%%%%%%%%%%%%%%%%%%%%%%%%%%%%%%%%%%%%%%%%%%%%%%%%%%%%%%%%%%%%%%%%%%%%%
	Let us start with a brief review of known results in flat space.
	The standard treatment in our language, as found in \cite{Parker} (and references therein) reads as follows. Consider the heat kernel for a free scalar of mass $m$ in $\mathbb{R}^{n-1}\times S^1$. Let us denote coordinates as $x^\m\in \mathbb{R}^{n-1}$ where $(\m=0,\ldots, n-2)$ and $y\equiv \theta R_0 \in S^1$. We have denoted the radius of the compact dimension by $R_0$ in order to avoid confusion with the scalar curvature which we denote by $R$. With this, the boundary conditions we need to impose are
	\be
	\phi(x,y)=\phi(x,y+2\pi R_0) = \phi(x,y+L).
	\ee
	To implement this periodicity, we can expand the fields in modes as
	\be
	\phi = \sum_k \dfrac{1}{\sqrt{L}} \phi_k(x) e^{i 2 \pi k y/L} =\dfrac{1}{\sqrt{2 \pi R_0}} \sum_k \phi_k(x) e^{i k y/ R_0}.
	\ee
	It should be remarked that whereas the dimension $[\phi]={n-2\over 2}$ (that is $1$ in $n=4$ dimensions), the dimension of $[\phi_k]={n-3\over 2}$ ($1/2$ in four dimensions).
	\par
	Let us take the simple example of a massive scalar field with a $\lambda \phi^4$ interaction in a four dimensional space where one of the coordinates is compactified on a circle. Expanding the scalar field as a background value and a perturbation, the quadratic piece in the perturbation reads
	\be
	S_2=-\frac{1}{2}\int d(vol)_4\, \phi(x,y)\left(\Box_4+M^2\right) \phi(x,y),
	\ee
	where $M^2 \equiv  m^2+\dfrac{\l}{2}\bp^2$ with $\pd_\m \bp=0$ .
	\par
	The {\em effective potential} is defined as the approximation to the effective action in which $\bar{\phi}$ is constant. This is the first term in an expansion of the background field in derivatives. Going back to the spacetime decomposition we can write the quadratic operator as 
	\bea
	S_2 &&=-\frac{1}{2} \int d^3x \, \sqrt{|\bg|^{(3)}} \int_0^L dy  \dfrac{1}{L} \sum_k \sum_l \phi_k(x) \left[\Box^{(3)} + \left(\dfrac{2 \pi k}{L}\right)^2 + M^2\right] \phi_{l} (x) e^{i (2 \pi /L) (k+l)y}= \nonumber \\
	\nonumber \\
	&& =- \frac{1}{2}\int d^3x \, \sqrt{|\bg|^{(3)}}  \sum_l \phi_l(x) \left[\Box^{(3)} + \left(\dfrac{2 \pi l}{L}\right)^2 + M^2\right] \phi_{-l} (x),
	\label{quad3d}
	\eea
	where we clearly see that the effect of integrating in the compact dimension is a shift in the effective mass of the scalar field. At this point, we are working with real scalar fields so we have $\phi_{-l} (x) = \phi_{l} (x)$.
	%%%%%%%%%%%%%%%%%%%%%%%%%%%%%%%%%%%%%%%%%%%%%%%%%%%%%%%%%%%%%%%%%%%%%%%%%%%%%%%%%%%%%%%%%%%%%%%%%%%%%%%%%%%%%%%%%%%%%%%%%
	\par
	Were it not for the fact that one of the dimensions is a circle, we would have that the effective action reads 
	\be
	W =- \int d^n x  \sqrt{|\bg|} \int_0^\infty \dfrac{d\t}{\t}(4\pi\t)^{- n/2}e^{-m^2 \t}=-\frac{V_n}{(4\pi)^{n/2}} \left(m^2\right)^{n/ 2}\Gamma\left(-n/2\right)
	\ee
	which is divergent for $n\in 2 \mathbb{N}$.
	Taking the precise case of \eqref{quad3d}, the  effective action corresponds  to $n-1$ dimensions really, owing to the fact that one of the spatial dimensions is compactified being thus equivalent to a Kaluza-Klein tower of momentum states. Using the effective mass of \eqref{quad3d} we get
	\be
	W=-{V_{n-1}\over (4\pi)^{{n-1\over 2}}}\,\Gamma\left({1-n\over 2}\right){1\over L^{n-1}}\sum_{l=-\infty}^\infty \left(M^2 L^2+ 4 l^2\pi^2\right)^{n-1\over 2}.\label{WM}
	\ee
	\par
	%This fact renders the value of the $\Gamma$ function finite. 
	This is then the effective potential in our case, including the quartic interaction in the effective potential approximation; that is, constant $\bp$. In the massless case, for $n=4$, we obtain
	\be
	\mathcal{E}_0=\frac{W}{V_3}=-\frac{\pi^2}{45L^3}.\label{Casi}\ee
	this result corresponds to the usual Casimir energy per unit volume computed in \cite{Parker}.  The remarkable fact is that it is negative definite, not the most natural thing to be  for an energy density.
	\par
	In the case of  $M\neq 0$, we focus in the summation of \eqref{WM} defining the sum
	%\be
	%S(M)\equiv\sum_{n=-\infty}^{\infty}\Big[ M^2\,L^2+n^2\pi^2\Big]^{(d-1)/2}=\left(M\,L\right)^{d-1}\left(1+2 \sum_{n=1}^\infty\left(1+\left({n\pi\over M L}\right)^2\right)^{d-1\over 2}\right)
	%\ee
	%We could use here the generalized binomial theorem
	%\be
	%(x+y)^\l=\sum_{k=0}^\infty {(\l)_k\over k!}\,x^{\l-k} y^k
	%\ee
	%(for  $|x| > |y|$). 
	% $(\l)_k$ is Pochhammer's symbol (falling factorial)
	%\be
	%(\l)_k\equiv \l(\l-1)\ldots(\l-k+1)
	%\ee
	%and $(\l)_0=1$.
	%By the same token
	%\be
	%\sum_{n=1}^\infty\left({n\pi\over M L}\right)^{2k}=\left({\pi\over M L}\right)^{2k}\zeta(-2k)
	%\ee
	%so that
	%\be
	%\sum_{n=1}^\infty\left(1+\left({n\pi\over M L}\right)^2\right)^{d-1\over 2}=\sum_{k=0}^\infty {\left({d-1\over 2}\right)_k\over k!}\left({\pi\over M L}\right)^{2k}\zeta(-2k)
	%\ee
	%Eventually we get our final expression for the sum
	%\be
	%S(M)=\left(M\,L\right)^{d-1}\left(1+2 \sum_{k=0}^\infty {\left({d-1\over 2}\right)_k\over k!}\left({\pi\over M L}\right)^{2k}\zeta(-2k)\right)
	%\ee
	
	%This expresion is really adequate  for $M L>> 1$ only, but we have included the correct expression for a massless field for consistency, that is, if $M=0$
	%\be
	%S(0)=2 \zeta\left(1-d\right)\label{S0}
	%\ee
	%\par
	%In fact, we could also start from
	\be
	S(M)\equiv\sum_{l=-\infty}^{\infty}\Big[ M^2\,L^2+4l^2\pi^2\Big]^{(n-1)/2}=\left( ML\right)^{n-1}+ 2  \sum_{l=1}^\infty (2l\pi)^{n-1} \Bigg[1+\left({ M L\over 2l\pi}\right)^2\Bigg]^{n-1\over 2}.
	\ee
Using the generalized binomial theorem\footnote{We have that \be
		(x+y)^\l=\sum_{k=0}^\infty {(\l)_k\over k!}\,x^{\l-k} y^k,
		\ee
		where we need  $|x| > |y|$ and  where $(\l)_k\equiv \l(\l-1)\ldots(\l-k+1)$ is the definition of Pochhammer's symbol (falling factorial) and $(\l)_0=1$.} and the definition of the zeta function the sum reads
	\be
	S(M)=\left( ML\right)^{n-1}+ 2  \, \sum_{k=0}^\infty {\left({n-1\over 2}\right)_k\over k!}\left({M L}\right)^{2k}\zeta(2k+1-n)(2\pi)^{n-1-2k}.
	\ee
	Let us note that the $k=0$ terms reproduces the previous massless case 
	\be
	S(0)=(2\pi)^{n-1} \zeta\left(1-n\right)\label{S0}.
	\ee
	To get the result for a complex scalar for Dirichlet boundary conditions at $y=0$ and $y=2\pi R_0$, we have to replace
	\be
	L\rightarrow 2 L.
	\ee
	We would like to emphasize that we have not attempted to compute the vacuum energy of the full flat space; rather our renormalization condition is precisely
	\be
	\lim_{L\rightarrow\infty} W_0=0
	\ee
	That is, we define the vacuum energy of the full flat space as zero and refer all other energies to it.
%%%%%%%%%%%%%%%%%%%%%%%%%%%%%%%%%%%%%%%%%%%%%%%%%%%%%%%%%%%%%%%%%%%%%%%%%%%%%%%%%%%%%%%%%%%%%%%%%%%%%%%%%%%%%%%
	\section{Vacuum energy induced in three-dimensional tori}
	%%%%%%%%%%%%%%%%%%%%%%%%%%%%%%%%%%%%%%%%%%%%%%%%%%%%%%%%%%%%%%%%%%%%%%%%%%%%%%%%%%%%%%%%%%%%%%%%%%%%%%%%%%%%%%%
	The purpose of this section is to study the vacuum energy of quantum field theory in a  background space-time of the form
	\be\label{product}
	\mathbb{F}_3\times \mathbb{R},
	\ee
	where $\mathbb{F}_3$ is a flat 3-manifold and $\mathbb{R}$ represents time. There are four-dimensional flat manifolds which  fail to be in this class, but we prefer to stick to \eqref{product}  for simplicity.
	These manifols have been completely classified by Joseph Wolf in \cite{Wolf}.
	\par
	Let us dwell in more detail in the particular case of $ \mathbb{F}^3 = \mathbb{T}^3= {\mathbb{R}^3 \over \Gamma}$ where $\Gamma$ is a three-dimensional lattice and the flat manifold corresponds to a general three-torus (computations on similar backgrounds have been carried out in \cite{odintsov3}). The mathematical  definition of a {\em lattice} \cite{Serre}  is the set of points in $\mathbb{R}^3$ of the form
	\be
	\Gamma\equiv \left\{ \mathbb{Z}\, \vec{e}_a\right\}.
	\ee
	The three dimensional vectors $\vec{e}_a$ (a=1\ldots 3) are the {\em generators} of the lattice. Accordingly, the {\em dual lattice} $\Gamma^*$ is the set of points $w\in \mathbb{R}^3$ such that, $w.v\in \mathbb{Z}$, for all points $v\in \Gamma$. 
	\iffalse
	Let us check when two lattices are equivalent.
	\be
	\Gamma^*\sim \Gamma
	\ee
	it is clear that in order for it to be so, we need that
	\be
	\vec{e}_a\,^*= L_a\,^b\, \vec{e}_b
	\ee
	where $L_a\,^b \in SL\left(3,\mathbb{Z}\right)$, in order for $L^{-1}$ to also lie in $ SL\left(3,\mathbb{Z}\right)$ (cf. Appendix \ref{B}) and so be able to express the basis $\vec{e}_b$ in terms of the $\vec{e}_a\,^*$ basis.
	\par
	Sometimes (like in string theory compactifications) we are interested in {\em self-dual} lattices
	\be
	\Gamma=\Gamma^*
	\ee
	\fi
	Now we can define the metric in this space as
	\be
	g_{ab}\equiv \vec{e}_a.\vec{e}_b\quad a,b=1\ldots 3,
	\ee
	which we will assume to be non-degenerate and positive definite. The dual lattice $\Gamma^*$ is generated by the vectors $\vec{e}\,^*_a$ such that
	\be
	\vec{e}\,^*_a. \vec{e}_b=\d_{ab}\label{ee}.
	\ee
	We shall define the {\em volume} of the lattice by $Vol(\Gamma)\equiv \det\,g_{ab}$ and dub the lattice as {\em unimodular} if $Vol(\Gamma)=1$.
	\par 
	In a 3-torus $\mathbb{T}^3 =\mathbb{R}^3/\Gamma$ points are identified under 
	\be
	x^i= x^i+\sum_{a=1}^{a=3} n^a\, 2\pi R_a\, e_a^i,
	\ee
	where the sub-index in $R_a$ indicates a different radius for each direction. 
	\iffalse
	The lattice is {\em integral} if
	\be
	\vec{v}.\vec{w}\in \mathbb{Z}
	\ee
	for all $\vec{v},\vec{w}\in \Gamma$. An integral lattice is {\em even} if
	\be
	\vec{v}.\vec{w}\in 2\mathbb{Z}
	\ee
	and {\em odd} otherwise.
	\par
	\fi
	We can now define some new coordinates using \eqref{ee}, live in circles, $z_a\equiv R_a \theta_a$, and are defined as
	\be
	z_a\equiv   \vec{x}\vec{e^*}_a=\vec{x}.\vec{e^*}_a+2\pi n_a R_a,
	\ee
	withe the periodicity property $ z_a=z_a+2\pi n_a R_a $.
	\iffalse
	\bea
	& x^i=x^j(e^*)^b_j (e^*)_b^i=z^a (e^*)_a^i\quad \Longrightarrow \quad dx^i= dz^b(e^*)_b^i=R_b\,d\theta^b\,(e^*)_b^i
	\eea
	We shall need later on the fact that
	\be
	{z_a\over 2\pi R_a}={z_a\over 2\pi R_a}+n_a
	\ee
	\fi
	In these coordinates, the corresponding spacetime metric will be
	\be
	ds^2\equiv f_{\m\n} \, dx^\m dx^\n = dt^2- \sum_{a,b=1}^3\,g^{ab}_* d z_a d z_b.
	%&&=dt^2-\sum_{a,b=1}^3\,(g^*)^{ab}\, R_a\, R_b\,  d \theta_a \,d \theta_b
	\label{metric}
	\ee
	\par 
	After describing the needed coordinates and metric for the precise spacetime, let us introduce an interacting quantum field in $\mathbb{F}_3\times\mathbb{R}$. The action we consider has the following form
	\be
	S = \int d^4 x \, \sqrt{|f|}  \bigg\{{1\over 2} f^{\m\n}\pd_\m \phi\pd_\n\phi-{1\over 2}\,m^2 \phi^2 - {1\over 4!}\l\phi^4 \bigg\}.
	\ee 
	where the metric has been defined in \eqref{metric}. Taking again the one-loop effective potential approximation, the  piece of the lagrangian quadratic in the quantum fields would read
	\be
	S_2 =\bar{S}+ \int d^4x \, \sqrt{|f|}\bigg\{{1\over 2} f^{\m\n}\pd_\m \phi\pd_\n\phi-{1\over 2}\bar{M}^2 \phi^2 \bigg\},
	\ee
	where the mass matrix is defined as $\bar{M}^2\equiv m^2+{1\over 2} \l\bp^2$. Notice that  we keep assuming that $\pd_\m\bp=0$. The heat equation reads
	\be
	{\pd \over \pd\t} K(x-x^\prime|\t)=-\left[\Box_E +\bar{M}^2\right]\,K(x-x^\prime| \t)\label{HK}
	\ee
	where $x=(t,z_a)$ as before and the $\Box_E$ operator stands for the euclidean\footnote{We are working with the mostly minus signature so that $\Box_E = - \dfrac{\pd^2}{\pd t^2}-\dfrac{\pd^2}{\pd z_a^2}$.} version of the Laplacian associated to the metric \eqref{metric}. Periodicity of the heat kernel in all the space of the $z$ coordinates is assured by construction as the solution is related to Riemann's theta function \cite{Mumford}
	\be
	\Theta\left(x-\xp|\Omega\right)\equiv \sum_{n\in \mathbb{Z}^g} e^{ i \pi n^2 \Omega + 2\pi i n.(x-\xp)}
	\ee
	where $x\in \mathbb{C}^g$ and $\Omega$ is a $g\times g$ complex matrix such that $\text{Im}\,\Omega > 0$. In our case we need $g=3$, see Appendix \eqref{D3} for more details.
	In particular, we make the following ansatz for the spatial part of the heat kernel
	\be\label{KO}
	K\left(z_a-z_a^\prime\middle|\Omega\t\right)=\Theta\left({z_a-z_a^\prime\over 2\pi R_a}\middle| \Omega\,\t\right).
	\ee
	Note that the Riemann theta function is periodic, see Appendix \eqref{R}. Taking the $\tau$ derivative we get
	%     \be
	%K\left(z_a-z_a^\prime\middle|\Omega\t\right)=\Theta\left({z_a-z_a^\prime\over 2\pi R_a}\middle| \Omega\,\t\right)\label{KO}
	%   \ee
	%now we put in the heat kernel equation \eqref{HK}, we get
	%  \bea
	%\Big[\pi i \sum_{\m\n} \Omega_{\m\n}  n_\m n_\n-M^2 \Big] K\left(z_a-z^\prime_a\middle|\t\right)=\Big[-\sum_{\m\n}\frac{n_\m n_\n}{R_\m R_\n}g_*^{\m\n} -M^2\Big]K\left(z_a-z^\prime_a\middle|\t\right)\nonumber\\
	%\eea
	% \bea
	%\Big[\pi i \sum_{\m\n} \Omega_{\m\n}  n_\m n_\n \Big] K\left(z_a-z^\prime_a\middle|\Omega\,\t\right)=\Big[-\frac{n_\m n_\n}{R_\m R_\n} \Big]K\left(z_a-z^\prime_a\middle|\Omega\,\t\right)
	%\eea
	\be
	{\pd \over \pd \t} K\left(z_a-z^\prime_a\middle|\Omega\,\t\right)=\Big[\pi i \sum_{\m\n} \Omega^{\m\n}  n_\m n_\n\Big] K\left(z_a-z^\prime_a\middle|\Omega\,\t\right),
	\ee
	which has to be equal to the spatial part of the heat kernel equation \eqref{HK}, namely, $g_*^{ab} \dfrac{\partial^2}{\partial z_a \partial z_b} K\left(z_a-z^\prime_a\middle|\Omega\,\t\right)$. This forces
	\be
	\Omega^{\m\n}={ i  \over \pi R_\m R_\n } g_*^{\m\n}, \label{Omega}
	\ee
	where the repeated indices do not indicate summation in this case. 
	%Please note that
	%\be
	%g_*^{ab} g^*_{bc}=\d^a_c.
	%\ee
	%To this we should append
	%   \be
	%  {1\over \sqrt{4\pi\t}} e^{-{(t-t^\prime)^2\over 4 \t}}
	% \ee
	\par
	We can finally write the total heat kernel as
	\be
	K\left(t-t^\prime, z_a-z_a^\prime\middle|\Omega \t\right)= {1\over \sqrt{4\pi\t}} e^{-{(t-t^\prime)^2\over 4\t}-\bar{M}^2\t}\,\Theta\left(\left.{z_a-\zp_a\over 2\pi R_a}\right|{i\,\t\,g_*^{\m\n}\over \pi R_\m R_\n}\right),
	\label{heatcompleto}
	\ee
	so that the effective potential energy reads
	\bea
	E_{0} = -\int\,{d\t\over \t}\,{1\over \sqrt{4\pi\t}}\,e^{ -\t\,\sum_{a b}  g_*^{ab}{n_a n_b\over R_a R_b}  -\bar{M}^2\t}=\Big[\sum_{a b}  g_*^{ab}{n_a n_b\over R_a R_b}+\bar{M}^2\Big]^{1/2}.\nonumber \\
	%&&=\Big[-i\pi\sum_{a b}  n_a\Omega_{ab} n_b+M^2\Big]^{1/2}
	\eea
	\subsection{Duality property}
	%%%%%%%%%%%%%%%%%%%%%%%%%%%%%%%%%%%%%%%%%%%%%%%%%%%%%%%%%%%%%%%%%%%%%%%%%%%%%%%%%%%%%%%%%%%%%%%%%%%%%%%%%
	After the preliminary computation of the induced vacuum energy on the three-dimensional tori, let us focus on the relation between this potential for radius $R$ and for radius $\widetilde{R} = \dfrac{l_s^2}{R}$ (where at this point $l_s$ is just a constant with dimensions of length), similar to the T-duality property in string theory (cf. \cite{Alvarez1994} and references therein). The key point in finding this relation is the modular property of the theta function, see Appendix \eqref{R}, which in the case of interest takes the form
	%In the particular case of $M_1$ %(cf. Appendix \ref{C})
	%\be
	%M_1\equiv \bpm 0&1\\-1&0\epm
	%\ee
	%the modular property reduces to
	\be
	\Theta\left(-\frac{1}{\t}\Omega^{-1} \mathbf{z}\middle|-\frac{1}{\t}\Omega^{-1}\right)=\sqrt{\det\left(-\Omega\t\right)}\,e^{\frac{i\pi}{\t} \ \mathbf{z} \Omega^{-1} \mathbf{z}}\,\Theta\left(\mathbf{z}|\Omega\t\right).
	\ee
	Using \eqref{Omega} we have that for our case
	\be
	\Omega^{-1}_{\m\n}=-i\pi R_\m R_\n g^*_{\m\n},
	\label{omegainv}
	\ee
	where again no summation is implicit.
	%\be
	%\Omega^{-1}_{\m\a} \Omega^{\a\n}=-i\pi R_\m R_\a g^*_{\m\a}{i\over \pi R_\a R_\n} g_*^{\a\n}={R_\m\over R_\n}\d_\m^\n=\d_\m^\n
	%\ee
	Taking the form of the spatial coordinates appearing in \eqref{heatcompleto} together with \eqref{omegainv}, we find the following relation 
	\bea
	\Theta\left({i\over 2\t}  R_\m\sum_\a  g^*_{\m\a}\left(z_\a-z^\prime_\a\right)\middle|{i\over \t}\pi R_\m R_\n g^*_{\m\n}\right) &=&
	\sqrt{\det\,\left(- \frac{i\t g_*^{\a\b}}{ \pi R_\a R_\b}  \right)}\, e^{\frac{1}{4\t}  \sum_{\r\s}\left( z^\r-z^{\prime\r}\right) g^*_{\r\s} \left(z^\s-z^{\prime\s}\right)}\times \nonumber\\
	&&  \times\Theta\left({z_\m-\zp_\m\over 2\pi R_\m}\middle|{i\,\t\,g_*^{\m\n}\over \pi R_\m R_\n}\right).
	\label{modprop}
	\eea
	\par
	This entails some relationship between theories compactified on $R_a$ and those compactified on ${1\over R_a}$, as we can use \eqref{modprop} to relate the spatial part of the heat kernel at each of the radius as
	%\bea
	%   K\left( {\widetilde{z}_\m-\widetilde{z}^\prime_\m\over 2\pi \widetilde{R_\m}} \left|\widetilde{\Omega}\widetilde{\t}\right)=\sqrt{\det\,\left(-{i\t g_*^{\a\b}\over  \pi R_\a R_\b}  \right)}\, e^{\frac{1}{4\t}  \sum_{\r\s}( z_\r-z_\r^\prime) g*_{\r\s} (z_\s-z_\s^\prime)}K \left( {z_a-\zp_a\over 2\pi R_a}\middle|\Omega\t\right)\right.\label{KK}\nonumber\\
	%\eea
	\bea
	&&K\left( \frac{\widetilde{z}_\m-\widetilde{z}^\prime_\m}{2\pi \widetilde{R_\m}} \middle|\widetilde{\Omega}\widetilde{\t}\right)= {1\over \sqrt{4\pi\widetilde{\t}}} e^{-M^2\widetilde{\t}}\,\Theta\left( {\widetilde{z}_\m-\widetilde{z}^\prime_\m\over 2\pi \widetilde{R_\m}} \middle|\widetilde{\Omega}\widetilde{\t}\right)=\nonumber\\
	&&=\sqrt{{\t\over \widetilde{\t}}} e^{-M^2(\widetilde{\t}-\t)}\sqrt{\det\,\left(-{i\t g_*^{\a\b}\over  \pi R_\a R_\b}  \right)}\, e^{\frac{1}{4\t}  \sum_{\r\s} ( z_\r-z_\r^\prime) g*_{\r\s} (z_\s-z_\s^\prime)}K\left( {z_\m-\zp_\m\over 2\pi R_\m}\middle|\Omega\t\right) \nonumber \\
	\label{KK}
	\eea
	where the tilde variables corresponding to the inverse radius\footnote{ There is another possibility given by  \bea
		&&\widetilde{\t}={\tau}, \quad \widetilde{R}_\m\equiv {\tau \over \pi  R_\m}, \quad \widetilde{g}^*_{\m\n}\equiv g^*_{\m\n}, \quad \widetilde{z}_\m=i \sum_\a g^*_{\m\a} z_\a.
		\eea
		Nevertheless it is not clear whether the $\tau$ dependence of $\tilde{R}$ interferes with its physical meaning.} read
	\be \label{RR}  \widetilde{\t}={\pi^2l_s^4\over \t}, \quad  \widetilde{R}_\m\equiv {l_s^2\over R_\m}, \quad {\widetilde{g}^*}_{\m\n}\equiv g^*_{\m\n}, \quad \widetilde{z}_\m={i\pi l_s^2 \over \t}\sum_\a g^*_{\m\a} z_\a. 
	\ee
	%\bea
	%&&\widetilde{\t}={\pi^2\over \t}\nonumber\\
	%&&\widetilde{R}_\m\equiv {1\over R_\m}\nonumber\\
	%&&\widetilde{g}_*^{\m\n}\equiv g^*_{\m\n}\nonumber\\
	%&&\widetilde{z}_\m={i\pi\over \t}\sum_\a g^*_{\m\a} z_\a
	% \eea
	Here $l_s$ is a (at this point arbitrary) length scale that is introduced to keep engineering dimensions right. Let us note that for $\t\in\mathbb{R}$ this relations map $z_a\in \mathbb{R}$ into $\widetilde{z}_a\in\mathbb{C}$, but the coordinates $z_a$ remain real for $\t$ imaginary. 
	\par
	%%%%%%%%%%%%%%%%%%%%%%%%%%%%%%%%%%%%%%%%%%%%%%%%%%%%%%%%%%%%%%%%%%%%%%%%%%%%%%%%%%%%%%%%%%%%%%%%%%%%%%%%%%%%%%%%%%%%%%%
	
	%Where however the first factor in the heat kernel, as well as the dependence on the total mass, are kept unchanged.
	%\be
	%{1\over \sqrt{4\pi\t}}e^{-{(t-t^\prime)^2\over 4\pi\t}-M^2\t}
	%\ee 
	Finally, we can compute the effective potential energy, which reads 
	\be
	E_0(R_a)=-\int {d\t\over \t}\tr\,K(\t)=-\int {d\t\over \t}{1\over \sqrt{4\pi\t}}e^{-M^2\t}\,\Theta\left(0\left|{i\t g_*^{ab}\over \pi R_a R_b}\right)\right.,
	\label{potentialR}\ee
	%where the spatial volume is given by
	%\be
	%V_3\equiv \int d^3 z\sqrt{g_*}.
	%\ee
	%where we have used the fact that
	%\be
	%\tr\,K(\t)={1\over\sqrt{4\pi\t}}e^{-M^2\t}\,\Theta\left(0\left|{i\t g_*^{ab}\over \pi R_a R_b}\right)\right. .\ee
	we can invert \eqref{potentialR} to write the theta function in terms of the effective potential as 
	\be
	{1\over \t\sqrt{4\pi\t}}\,\Theta\left(0\left|{i\t g_*^{ab}\over \pi R_a R_b}\right)\right.=-{1\over 2\pi i}\int_C d \mu^2 E_0(R_a) e^{\mu^2 \t}.
	\ee
	the circuit $C$ is the one corresponding to $Re\,\mu^2=c > 0$ in the complex $\mu^2$ plane ($c$ being an arbitrary positive constant).
	\par 
	In a similar way, we can compute the potential energy $\widetilde{E_0}$ corresponding to $\widetilde{R_a}$, which is itself a function of $R_a$ and $\tau$. This potential  energy then is going to depend on the normal radius $R_a$ and we can write it as a function of $E_0(R_a)$ using \eqref{KK} as 
	%\bea
	%\tilde{V}(R_a)&&=-\int {d\widetilde{\t}\over \widetilde{\t}}\tr\,K(\widetilde{\t})
	%\eea
	%nevertheless for the trace of the space-like piece of the heat kernel, \eqref{KK}, we find that
	% \be
	% \tr\,\widetilde{K}\left(\widetilde{ \t}\right)=e^{\pi i\over 4}\left({\t\over \pi}\right)^{3/2} {\sqrt{g_*}\over R_1 R_2 R_3 } \tr\,K(\t)
	% \ee
	%therefore
	\bea
	\widetilde{E_0}(R_a)&&=-\int {d\widetilde{\t}\over \widetilde{\t}}\tr\,K(\widetilde{\t})=-\int {d\widetilde{\t}\over \widetilde{\t}}\sqrt{{\t\over \widetilde{\t}}} e^{-M^2(\widetilde{\t}-\t)}e^{\pi i\over 4}\left({\t\over \pi}\right)^{3/2} {\sqrt{g_*}\over R_1 R_2 R_3 } \tr\,K(\t)=\nonumber\\
	&&=\int d\t \frac{\t}{2i\pi^2 l_s^2}e^{\pi i\over 4}\left({\t\over \pi}\right)^{3/2} {\sqrt{g_*}\over R_1 R_2 R_3 } e^{-M^2\pi^2 l_s^4/\t}\int_C d \m^2 E_0(R_a) e^{\m^2 \t}.
	\eea
	where we have used \eqref{RR}. 
	This non-local integral relationship between the potential and its dual is at variance  with the situation in string theory (see e.g. \cite{Alvarez1994} and references therein), where the relationship between the effective potentials for dual tori is much simpler (they are actually proportional).

	%%%%%%%%%%%%%%%%%%%%%%%%%%%%%%%%%%%%%%%%%%%%%%%%%%%%%%%%%%%%%%%%%
	%%%%%%%%%%%%%%%%%%%%%%%%%%%%%%%%%%%%%%%%%%%%%%%%%%%%%%%%%%%%%%%%%%%%%%%%%%%%%%%%%%%%%%%%%%%%%%%%%%%%%%%%%%%%%%%
	\section{The effect of dynamical gravity on the vacuum energy}
	%%%%%%%%%%%%%%%%%%%%%%%%%%%%%%%%%%%%%%%%%%%%%%%%%%%%%%%%%%%%%%%%%%%%%%%%%%%%%%%%%%%%%%%%%%%%%%%%%%%%%%%%%%%%%%%
	Let us now turn to the study of another aspect of vacuum energy, namely, the quantum gravity corrections to the Casimir effect (cf. \cite{WQG} and references therein). We aim to study the possible changes introduced by graviton fluctuations. Once dynamical gravity is considered, there is no ambiguity related to the energy-momentum tensor and the effective action retains all of the gauge invariance. 
	\par
	%It is not clear whether in quantum gravity topology changing amplitudes are allowed \cite{Alvarez2020}. At any rate, we are going to assume in this paper that this is not the case and that the topology remains the same as the background topology.
	In order to analyze the changes in the Casimir energy brought by dynamical gravitons, we start with the following simple action 
	\be
	S=\int \sqrt{|g|} \, d^4 x \, \bigg\{-{1\over 2 \kappa^2} R+ \dfrac 1 2 g^{\m\n} \pd_\m \Phi \pd_\n \Phi- \dfrac 1 2 m^2\,\Phi^2-{\l\over 4!}\Phi^4\bigg\}.
	\ee
	We are going to work on a manifold of the form, $M_4 \equiv M_{3}\times S^1$, where $M_3$ represents an arbitrary  three-dimensional manifold with Minkowskian signature and the remaining  spatial dimension is compactified on a circle. 
	In order to compute the one-loop effective action and the effective potential, we use the background field technique \cite{DeWitt}. We expand the fields in their background value and a perturbation as 
	\bea 
	&& g_{\m\n}\equiv \bg_{\m\n}+\kappa h_{\m\n}, \nonumber\\
	&&  \Phi\equiv \bp+\phi.
	\label{backgroundexp}
	\eea
	\par
	Let us note that in order to be able to compare with the usual Casimir effect in a non-dynamical background, we take the following form of the background metric
	\be
	\bg_{\m\n} dx^\m dx^\n= \sum_{n=0}^2 \bg_{\a\b}(x) dx^\a dx^\b + dy^2,
	\label{defmetrica}
	\ee
	where $dy^2=R_0^2 d\theta^2$. It is important to notice at this point that  we are giving up some of the background gauge invariance. Instead of $\text{Diff}(M_4)$ we will have $\text{Diff}(M_3)\times SO(2)$ with linear generators
	\be
	\xi=\sum_{i=0,1,2}\xi^i(x){\pd\over \pd x^i}+{\pd\over \pd y}.
	\ee
	this means that we are neglecting certain quantum fluctuations to keep our background metric form-invariant. Nevertheless, we will stick to this type of backgrounds to make the computations physically sensible.
	\par

	With the expansion \eqref{backgroundexp} and after gauge fixing, the quadratic piece of the action takes the form
	\be
	S_{2+gf} =\frac{1}{2} \int \, \sqrt{|\bg|} \, d^4x \, \Phi^A \Delta_{AB} \Phi^B,
	\ee
	where we have defined the generalized field 
	\be
	\Phi^A\equiv\begin{pmatrix}
		h^{\a\b}\\\phi
	\end{pmatrix},
	\ee
	and the operator has the symbolic form given by
	\bea
	\Delta_{AB}=-g_{AB}\bar{\Box}  + Y_{AB}.
	\eea
	The details of the computation can be found in Appendix \eqref{S} (cf. also \cite{Gilkey}). In a previous paper \cite{AlvarezF}, we studied the two possible viewpoints that can be considered when renormalizing  Kaluza-Klein theories. The first one consists of renormalizing the higher dimensional theory first and expanding the resulting higher dimensional effective theory (including counterterms) in harmonics afterward. The other viewpoint consists of first expanding in harmonics the classical theory and renormalizing the resulting four-dimensional theory. The two viewpoints are in agreement for free theories \cite{Duff}, but not anymore when interactions are considered.
	\par 
	We shall stick here to the lower dimensional point of view, that is, the later alternative. We expand the fields in modes as 
	\be
	\Phi^A = \sum_k \Phi^A_k (x) e^{i k 2 \pi y /L},
	\label{modexp}
	\ee
	where $L= 2 \pi R_0$. We can integrate the periodic coordinate and get
	\be
	S_{2+gf} =\dfrac 1 2  \int \, d^3x \, \sqrt{|\bg|^{(3)}}  \, \sum_k \left( \Phi^A_k \Delta_{AB}^k \Phi^B_k \right),
	\ee
	where we have used \eqref{defmetrica} and $\Phi^A_k= \Phi^A_{-k}$.
	\par
	For the Casimir energy, we need to compute the finite part of the effective action. In order to do that, we are going to separate the contribution coming from the compact dimension, that is, the mode number dependence, as
	\bea
	\Delta_{AB}^k=-g_{AB}\bar{\Box}^{(3)}- \left(\frac{2k\pi}{L} \right)^2 g_{AB}+ Y_{AB}
	\eea
	Now, we know that using the heat kernel method the effective action reads
	\be
	W = - \int d^3x \, \sqrt{|\bg|^{(3)}}\, \text{Tr} \left\lbrace \int \dfrac{d \tau}{\tau} \, \sum_k K_k (x,x',\tau)\right\rbrace,
	\label{heat1}
	\ee
	with 
	\be
	K_k(x,x',\tau) = \dfrac{1}{(4\pi \tau)^{n/2}} e^{-M^k_{AB}\tau}  \sum_p a_p(\Delta_{AB}^k) \tau^p.
	\ee
	Note that we have defined the `mass matrix' as the part containing the induced masses coming from the compactification of the fourth dimension. In Appendix \eqref{D1} we show the equivalence between different ways of treating the mass term. In this case, we have
	\be
	M^k_{AB}=\left(\frac{2k\pi}{L} \right)^2 g_{AB}.
	\ee
	It is a fact that given the simple form of the matrix, it is possible to keep it in the exponential and treat it exactly without having to use the small proper time expansion. Nevertheless, the rest of the operator cannot be treated exactly so that we use the small proper time approximation for the remaining operator
	\bea
	\Delta_{AB}^k=  -g_{AB}\bar{\Box}^{(3)}+ Y_{AB}.
	\eea
	\par
	Integrating \eqref{heat1} over  $\tau$  yields
	\bea W = - \int d^3 x \, \sqrt{|\bg|^{(3)}} \, \dfrac{1}{(4\pi)^{3/2}} \mbox{tr}\sum_{k=-\infty}^{\infty}\sum_{p=0}^{\infty}\left({a_p}\right)^A_B \left[\left( {M^k}\right)^{3/2-p}\right]^{B}_{A} \, \Gamma\left(p-\frac{3}{2}\right).\eea
	we see that we need to multiply the matrix of the heat kernel coefficients with the mass matrix. Before going on, let us note that the mass matrix has a very simple form when we raise one of the generalized indices $A$, this is done using the internal metric defined in Appendix \eqref{S}, namely, 
	\be
	\left( {M^k}\right)^A_B = \left(\frac{2k\pi}{L} \right)^2 \delta^A_B,
	\ee
	so that any power of this matrix equals the identity matrix. Taking this into account and taking the sum of the first three heat kernel coefficients we get
	\bea W&&=- \int d^3 x \, \sqrt{|\bg|^{(3)}} \, \dfrac{1}{(4\pi)^{3/2}} \mbox{tr}\sum_{k=-\infty}^{\infty}\Big[ \left( \dfrac{2 k \pi}{L}\right)^3 \Gamma \left(-\dfrac{3}{2} \right)a_0(\Delta)^A_A + \left( \dfrac{2 k \pi}{L}\right) \Gamma\left(-\dfrac{1}{2} \right) a_1(\Delta)^A_A \nonumber \\
	&&+ \Gamma \left(\dfrac{1}{2} \right) \left(\dfrac{L}{2 k \pi }\right)a_2(\Delta)^A_A \Big]\eea
	In order to extract the finite part of the mode sums, we use here the zeta function regularization given near $s=1$ by 
	\be \zeta(s)= \sum_{n=0}^\infty \, \dfrac{1}{n^s} = \frac{1}{s-1}+\g_E-\g_1(s-1)+\mathcal{O}(s-1)^2,\ee
	so that we take the $\g_E$ as the finite part of $\zeta(1)$ (the details of the regularization can be found in Appendix (\ref{E}). We also need the values of $\zeta(-1)$ and $\zeta(-3)$ which are well-known. 
	With all of this, the effective action finally reads
	\bea \label{WW}&&W= -\int d^3 x \, \sqrt{|\bg|^{(3)}} \,  \Bigg\{\frac{\pi^2}{15L^3}+\frac{1}{24L}\Big[2m^2+\left(10m^2\k^2+\l\right)\bp^2+\frac{5}{6}\k^2\l\bp^4-11\k^2\bn_\n\bp\bn^\n\bp\Big]+\nonumber\\
	&&+\frac{L}{4 \pi^2} \g_E\Big[\frac{m^4}{4}+\left(-\frac{13}{6}m^4\k^2+\frac{m^2}{4}\l\right)\bp^2+\left(-\frac{57}{40}m^4\k^4-\frac{55}{72}m^2\k^2\l+\frac{\l^2}{16}\right)\bp^4-\nonumber\\
	&&-\frac{\k^2\l}{80}\left(19m^2\k^2+5\l\right)\bp^6-\frac{19}{1920}\k^4\l^2\bp^8-\k^2\left(2m^2+\frac{1}{3}\l\bp^2\right)\bp\bar{\Box}\bp-\frac{11}{12}m^2\k^2\bn_\m\bp\bn^\m\bp+\nonumber\\
	&&+\left(\frac{57}{40}m^2\k^4-\frac{11}{24}\k^2\l\right)\bp^2\bn_\m\bp\bn^\m\bp+\frac{19}{160}\k^4\l\bp^4\bn_\m\bp\bn^\m\bp+\k^2\bar{\Box}\bp\bar{\Box}\bp+\frac{203}{80}\k^2\bn_\m\bp\bn^\m\bp\bn_\n\bp\bn^\n\bp\Big]\Bigg\} \nonumber \\
	\eea
	%	Let us note that in the simplest case $m=\l=0$ and $n=4$, using the scalar equation of motion $\bar{\Box}\bp=0$ we reproduce 't Hooft and Veltman's result \cite{tHooft}. \textcolor{red}{Yo quitaria esto porque en realidad lo que se obtiene es thooft y veltmann antes de sumar en los modos}
	\par
	Taking $\kappa=0$ in our result does not yield directly the purely scalar part of it. Instead, we get the sum of the contributions of the scalar field in a fixed background and the purely gravitational part. This result can be understood by noticing that the $\kappa \rightarrow 0$ limit is equivalent to the decoupling limit of gravity. In this limit,  there is no interaction between the gravitons and the scalar field and the result is the independent sum of the contributions of the different fields.  %Nevertheless, if we take $\kappa \rightarrow 0$ in the result, all the terms coming from the graviton-scalar interaction go to zero and we are left with the sum of pure dynamical gravity plus the minimally coupled scalar field in a gravitational background. We can see this comparing the result for the scalar field in a gravitational background. 
	
	For massless scalars, in flat space, the well-known result is given in equation \eqref{Casi}.
	%\bea W&&=-\frac{V_4}{2\pi R^4}(4\pi)^{-3/2}\Gamma\left(-\frac{3}{2}\right)2\zeta(-3) = -\frac{V_4\pi^2}{720 r^4} \eea
	Taking  the $1/L^3$ contribution in \eqref{WW}, and  splitting  it as the sum of the pure gravitational piece (which contains the contribution of the ghost lagrangian), and the piece coming from the scalar field in the gravitational background, we have
	\be
	\mathcal{E}_0=-\frac{\pi^2}{15 L^3}= \frac{1}{V_3}\left(W_g+W_\phi\right)=  -\frac{2\pi^2}{45 L^3}-\frac{\pi^2}{45 L^3}.
	\label{Escalar}
	\ee
	The purely scalar part matches the Casimir energy found in \eqref{Casi}. It is worth to highlight the fact that the contribution of gravitons to the vacuum energy is exactly {\em twice} the one of a single scalar. This is what happens in flat spacetime (for an incomplete list of references see \cite{Nima,Fornal1,Fornal2,Ibanez:2017kvh,shiu,alvaro}), but we see here that it stays true even in our quite general spacetime backgrounds.
	%With the effective action 
	%\bea W= -\int d^3 x \, \sqrt{-g^{(3)}} \mathcal{L}\eea
	We can now compute the one-loop ``energy-momentum tensor" given by
	\bea &&T^{\m\n}=\frac{2}{\sqrt{|\bg|^{(3)}}}\frac{\d W}{\d \bg_{\m\n}}=-\frac{\pi^2}{15L^3}\bg^{\m\n}-\frac{1}{12L}\Bigg\{\Big[m^2+\left(5m^2\k^2+\frac{\l}{2}\right)\bp^2+\frac{5}{12}\k^2\l\bp^4-\frac{11}{2}\k^2\bn_\l\bp\bn^\l\bp\Big]\bg^{\m\n}+\nonumber\\
	&&+11\k^2\bn^\m\bp\bn^\n\bp\Bigg\}-\frac{L}{8\pi^2} \g\Bigg\{\Big[\frac{m^4}{2}+\frac{1}{6}\left(-26m^4\k^2+3m^2\l\right)\bp^2+\left(-\frac{57}{20}m^4\k^4-\frac{55}{36}m^2\k^2\l+\frac{\l^2}{8}\right)\bp^4-\nonumber\\
	&&-\frac{\k^2\l}{40}\left(19m^2\k^2+5\l\right)\bp^6-\frac{19}{960}\k^4\l^2\bp^8+\frac{13}{6}m^2\k^2\bn_\l\bp\bn^\l\bp+\frac{1}{60}\left(171m^2\k^4-55\k^2\l\right)\bp^2\bn_\l\bp\bn^\l\bp+\nonumber\\
	&&+\frac{19}{80}\k^4\l\bp^4\bn_\l\bp\bn^\l\bp+2\k^2\bar{\Box}\bp\bar{\Box}\bp+\frac{203}{40}\k^2\bn_\r\bp\bn^\r\bp\bn_\s\bp\bn^\s\bp\Big]\bg^{\m\n}-\frac{13}{3}m^2\k^2\bn^\m\bp\bn^\n\bp-\nonumber\\
	&&-\frac{1}{30}\left(171m^2\k^4-55\k^2\l\right)\bp^2\bn^\m\bp\bn^\n\bp-\frac{19}{40}\k^4\l\bp^4\bn^\m\bp\bn^\n\bp-\frac{203}{10}\k^2\bn^\m\bp\bn^\l\bp\bn_\l\bp\bn^\n\bp+\nonumber\\&&+8\k^2\bn^\m\bar{\Box}\bp\bn^\n\bp\Bigg\}\label{T}\eea
	Taking $\bp$ constant, the result for  massless scalars with no interaction reduces to
	\bea &&T^{\m\n}=-\frac{\pi^2}{15L^3}\bg^{\m\n},
	\eea
	which is in agreement with the classical references \cite{Brown, Fulling} to the extent that they can be compared. They are mainly interested in the parallel plates situation; whereas we are computing the change in vacuum energy due to  compactification in a circle.
	\section{Dynamical transverse gravity}
	%%%%%%%%%%%%%%%%%%%%%%%%%%%%%%%%%%%%%%%%%%%%%%%%%%%%%%%%%%%%%%%%%%%%%%%%%%%%%%%%%%%%%%%%%%%%%%%%%%%%%%%%%%%%%%
	When we functionally integrate over unimodular metrics {\em only} (which is of course not the same thing as GR in the gauge $|\bg|=1$) then the background-field-independent term in the effective action does not couple at all to the graviton (just because $\sqrt{|\bg|}=1$). Nevertheless in this case a curious thing happens. Namely, the invariance under transverse diffeomorphisms (TDiff) with generators that obey
	\be
	\pd_\m \xi^\m=0,
	\ee
	is not enough to imply conservation of the energy-momentum tensor corresponding to the background fields, but only guarantees the existence of some spacetime function $T(x)$ such that
	\be
	\nabla_\a{\pd S\over \pd g_{\a\b}}=\pd^\b T(x)
	\ee
	As is well known, Bianchi identities allow now for an {\em arbitrary} cosmological constant, which appears here as an integration constant in the background equations of motion. But the role of this integration constant seems to be somewhat mysterious in the sense that it does not couple with the graviton at all. It could even be that this means that only the zero value for this constant is fully consistent.
	\par
	%%%%%%%%%%%%%%%%%%%%%%%%%%%%%%%%%%%%%%%%%%%%%%%%%%%%%%%%%%%%%%%%%%%%%%%%%%%%%%%%%%%%%%%%%%%%%%%%%%%%%%%%%%%%%%%%%%%%%%%%%
	In this section we want to perform the same computation we did for GR but restricted to the unimodular theory. The unimodular action corresponding to a scalar field minimally coupled to the gravitational field can be written in terms of an unconstrained metric $\hat{g}_{\m\n}$ as
	\be
	S_{UG}=-{1\over 2 \kappa^2}\int d^n x\, |\hat{g}|^{1\over n}\,\left(\hat{R}+2\Lambda+{(n-1)(n-2)\over 4 n^2  }{\hat{g}^{\m\n} \pd_\m \hat{g}\pd_\n \hat{g}\over \hat{g}^2} - \kappa^2 \ \hat{g}^{\m\n}\,\pd_\m\phi\pd_\n\phi\right),
	\label{UGact}
	\ee
	where we have written the original unimodular metric as ${g}_{\mu\nu}= \hat{g}^{-1/n} \hat{g}_{\m\n}$. This action however has a complicated symmetry sector because of the artificial Weyl invariance that we have introduced when writing the theory in terms of an unconstrained metric (the theory is invariant under $\hat{g}_{\m\n} \rightarrow \Omega^2 \hat{g}_{\m\n}$). 
	In order to be able to carry out the computation, we shall employ a trick first devised in \cite{Alvarez2008}. Let us go through their arguments to introduce the framework we will use. 
	\par
	We can first generalize the unimodular action   by incorporating some arbitrary functions of the determinant of the metric in front of the invariant measure
	\be\label{Faedo}
	S=-{1\over 2 \kappa^2}\int d^n x\,\sqrt{|\hat{g}|} \left(f(\hat{g})  \hat{R} + 2 \Lambda f_\Lambda(\hat{g})+f_\varphi(\hat{g}) \hat{g}^{\m\n}\pd_\m \hat{g}\pd_\n \hat{g}  - \kappa^2 \, f_\phi(\hat{g})\hat{g}^{\m\n}\pd_\m\phi\pd_\n\phi\right).
	\ee
	in this way, we now have the most general transverse diffeomorphism ({\em TDiff}) invariant action, and the unimodular action \eqref{UGact} is then a particular case of \eqref{Faedo} for the following values of the functions
	\bea
	&& f(x)=x^{2-n\over 2n},\nonumber\\
	&&f_\varphi(x)=\frac{(n-1)(n-2)}{2n^2}\,x^{2-5n\over 2n},\nonumber\\
	&&f_\phi(x)=x^{{2-n\over 2n}},\nonumber\\
	&&f_\Lambda(x)=x^{2-n\over 2n}.
	\label{unimtransf}
	\eea
	\par
	The action \eqref{Faedo} is only invariant under the diffeomorphisms that leave the determinant unchanged. Nevertheless, we can now introduce a {\em compensator field} $C(x)$ such that
	\be
	\hat{ \s}(x)\equiv \hat{g}\, C^2(x),
	\ee
	transforms as a true scalar, and then, we restore full diffeomorphism invariance (at the cost of introducing a new degree of freedom). The {\em TDiff} invariant action corresponds to the {\em unitary gauge} $C=1$.   The generalized action then reads
	\bea
	S&=&-{1\over 2 \kappa^2} \, \int d^n x \sqrt{|\hat{g}|}\left( f(\hat{ \s})  \hat{R} + 2 \Lambda f_\Lambda(\hat{ \s})+f_\varphi(\hat{ \s}) \hat{g}^{\m\n}\pd_\m \hat{ \s}\pd_\n \hat{ \s}-\k^2f_\phi(\hat{ \s}) \hat{g}^{\m\n}\pd_\m\phi\pd_\n\phi\right).\nonumber\\
	\eea
	\par
	As a final step, we want to change to the Einstein frame so that the kinetic term of the graviton takes the canonical form. We start by performing a Weyl rescaling
	\bea
	&g_{\m\n}\equiv \Omega^2 \hat{g}_{\m\n},\nonumber\\
	&\s\equiv  \Omega^{2n}\hat{ \s},
	\eea
	where the conformal factor $\Omega$ is such that $\Omega^{n-2}=f(\hat{ \s})$. In this way, the gravitational piece of the action is written in Einstein's frame, so that we have
	\be
	\sqrt{|\hat{g}|}f(\hat{ \s})\hat{R}=\sqrt{|g|} R+\ldots.
	\ee
	after this Weyl transformation, the action transforms into
	\bea
	S&=&-{1\over 2 \kappa^2}\int d^n x \sqrt{|g|} \Big[R+ 2\Lambda F_\Lambda(\Omega)-\k^2 f_\phi (f^{-1}\left(\Omega^{n-2}\right))\Omega^{2-n}\, g^{\m\n}\pd_\m\phi\pd_\n\phi \Big]+\nonumber\\
	&+&{1\over 2 \kappa^2}\int d^n x\, \sqrt{|g|}\left[{2(n-1)(n-2)\over \Omega^2}-\Omega^{2-n} f_\varphi( f^{-1}\left(\Omega^{n-2}\right)) \left({\pd f^{-1}\left(\Omega^{n-2}\right)\over \pd \Omega}\right)^2 \right]\,g^{\m\n}\pd_\m\Omega\pd_\n\Omega, \nonumber \\
	\eea
	where we have defined $F_\Lambda(\Omega)\equiv \Omega^n f_\Lambda (f^{-1}\left(\Omega^{n-2}\right))$. 
	We can now make one final redefinition given by 
	\be
	\left[{2(n-1)(n-2)\over \Omega^2}-\Omega^{2-n} f_\varphi( f^{-1}\left(\Omega^{n-2}\right)) \left({\pd f^{-1}\left(\Omega^{n-2}\right)\over \pd \Omega}\right)^2 \right]\,g^{\m\n}\pd_\m\Omega\pd_\n\Omega \equiv \k^2 g^{\m\n} \pd_\m \varphi \pd_\n\varphi.
	\label{redef}
	\ee
	After all these steps we finally arrive at a quite simple action for gravity coupled to two scalar degrees of freedom, one of them with a non-minimal coupling, which reads
	\bea
	S&=&-{1\over 2 \kappa^2}\int d^n x \sqrt{|g|} \left[R+2 \Lambda F_\Lambda(\varphi)-\k^2 g^{\m\n} \pd_\m \varphi \pd_\n\varphi-\k^2 F_{\phi}(\varphi) \, g^{\m\n}\pd_\m\phi\pd_\n\phi \right].
	\eea
	In this formula, we have also defined $F_{\phi}(\varphi) = f_\phi (f^{-1}\left(\Omega^{n-2}\right))\Omega^{2-n}$. 
	\par
	As an important remark, let us mention however that the preceding set of transformations are not strictly valid in the unimodular case because \eqref{redef} vanishes when particularizing it for \eqref{unimtransf}.  This means that there is no way of writing $\Omega (\varphi)$ if the kinetic term vanishes for the new field. This leads to the non-invertibility of the Weyl transformations so that we cannot go back to the Jordan frame. In other words, there is no way to implement Einstein's frame in unimodular gravity via a Weyl transformation.
	Nonetheless, there is some  evidence based upon the results in \cite{Alvarez2008}, that computing for general $f(x)$  and particularizing at the end to the value $f(x)\rightarrow x^{2-n\over 2 n}$ one gets the correct result, at least for the divergent piece of the effective action. In particular, it was shown there that whenever
	\be
	2(n-1) f^\prime(x)^2-(n-2)f(x)f_\varphi(x)=0,
	\ee
	the theory is on-shell one-loop finite \cite{Alvarez2008}. Unimodular Gravity corresponds to
	\be
	f_\varphi(x)={(n-1)(n-2)\over 2 n^2} x^{2-5n \over 2n},
	\ee
	that is, it saturates this equality. It is quite remarkable that the only transverse theories which are on-shell one-loop finite are precisely Einstein's general relativity and Unimodular Gravity.
	\par
	In this section, we carry on with the computation for a general transverse theory. % \bea
	%&& S=-{1\over 2 \kappa^2}\int d^n x \sqrt{g} \left(R+2 \Lambda F_\l(\Omega)\right)+{1\over 2}g^{\m\n} \pd_\m \varphi \pd_\n\varphi\nonumber\\
	%&&+{1\over 2}\int d^n x\, \sqrt{g}\left(f_\phi \circ f^{-1}\right)\left(\Omega^{n-2}\right) \left( \Omega^{2-n}\, g^{\m\n}\pd_\m\phi\pd_\n\phi  - \Omega^{-n} \dfrac{m^2}{2} \phi^2- \Omega^{-n} \dfrac{\l}{4!} \phi^4 \right) \eea
	%We can define 
	%\be
	% F_{\phi}(\varphi) = \left(f_\phi \circ f^{-1}\right)\left(\Omega^{n-2}\right)  \Omega^{2-n},
	%\quad \quad  F_{\phi,m}(\varphi) = \left(f_\phi \circ f^{-1}\right)\left(\Omega^{n-2}\right)  \Omega^{-n} \ee
	%so that 
	%\bea
	%&& S=-{1\over 2 \kappa^2} \int d^n x \sqrt{g} \left(R+2 \Lambda F_\l(\varphi)\right)+{1\over 2 \kappa^2}g^{\m\n} \pd_\m \varphi \pd_\n\varphi\nonumber\\
	%&&+\int d^n x\, \sqrt{g} \, \left( {1\over 2} F_{\phi}(\varphi) \, g^{\m\n}\pd_\m\phi\pd_\n\phi  - F_{\phi,m}(\varphi) \, \left(  \dfrac{m^2}{2} \phi^2+\dfrac{\l}{4!} \phi^4 \right) \right) \eea
	% This action is already in a form in which standard background field techniques can be applied.
	In order to make the computation feasible, we will expand the scalar fields around constant backgrounds  (so that the kinetic term of the real scalar field is just $- F_{\phi}(\bar{\varphi})\phi \bar{\Box} \phi$ and the non-diagonal terms with derivatives vanish). The scalar equation of motion implies that  $\bar{\phi}$ constant is  a solution only in the massless and non-self-interacting case. This is the reason why we take this simple example instead of the massive interacting scalar field of the previous section.
	\par 
	Let us start with the computation of this simple model. We have two scalar fields with constant backgrounds plus the graviton. Taking the quadratic piece after the expansion \eqref{backgroundexp}, together with the gauge fixing action, we have
	\be
	S_{2+gf} = \frac{1}{2}\int \, \sqrt{|\bg|} \, d^4x \, \Phi^A \Delta_{AB} \Phi^B,
	\ee
	where the generalized field is now
	\be
	\Phi^A\equiv\begin{pmatrix}
		h^{\a\b}\\\phi \\ \varphi
	\end{pmatrix}.
	\ee
	Again, the operator has the symbolic form
	\bea
	\Delta_{AB}=-\begin{pmatrix}
		C_{\a\b\m\n} &0&0\\0&F_\phi(\varphi)  &0 \\ 0& 0& 1 \end{pmatrix} \bar{\Box} + Y_{AB}.
	\eea
	The details of the computation can be found in Appendix \ref{S}.
	Performing the same mode expansion as before \eqref{modexp} we then have
	\be
	S_{2+gf} =\dfrac 1 2  \int \, d^3x \, \sqrt{|\bg|^{(3)}}  \, \sum_k \left( \Phi^A_k \Delta_{AB}^k \Phi^B_k \right),
	\ee
	where we can separate
	\iffalse
	\textcolor{blue}{JA: REPETIDO 	We expand the fields in modes as 
		\be
		\Phi^A = \sum_n \Phi^A_n (x) e^{i n 2 \pi y /L}	
		\ee
		so that we can integrate the periodic coordinate and get
		\be
		S_{2+gf} =\dfrac 1 2  \int \, d^3x \, \sqrt{\bar{\hat{g}}}  \, \sum_n \left( \Phi^A_n \Delta_{AB}^n \Phi^B_n \right),
		\ee
		where we have used \eqref{defmetrica} and $\Phi^A_n= \Phi^A_{-n}$.
		FIN JA}
	\par
	\fi
	%For the Casimir energy, we need to compute the finite part of the effective action. In order to do that, we are going to separate
	the contribution coming from the compact dimension and define 
	\bea
	\Delta_{AB}^k=-\begin{pmatrix}
		C_{\a\b\m\n} &0&0\\0& F_\phi(\varphi)  &0 \\ 0& 0& 1 \end{pmatrix}\bar{\Box}^{(3)}-  
	\left(\dfrac{2 k \pi}{L} \right)^2
	\begin{pmatrix}
		C_{\a\b\m\n} &0&0\\0& F_\phi(\varphi) &0 \\ 0& 0& 1 \end{pmatrix}+ Y_{AB}.\nonumber \\
	\eea
	Taking the same definitions of the previous section we then have
	\bea W = - \int d^3 x \, \sqrt{|\bg|^{(3)}} \, \dfrac{1}{(4\pi)^{3/2}} \mbox{tr}\sum_{k=-\infty}^{\infty}\sum_{p=0}^{\infty}\left({a_p}\right)^A_B \left[\left( {M^k}\right)^{3/2-p}\right]^{B}_{A} \, \Gamma\left(p-\frac{3}{2}\right).\eea
	We need to compute the heat kernel coefficients of the operator that we obtain when we subtract the part of the masses involving the mode number. 
	Again, the mass matrix has a very simple form when we raise one of the generalized indices $A$, namely, 
	\be
	\left( {M^k}\right)^A_B = \left(\dfrac{2 k \pi}{L} \right)^2 \delta^A_B,
	\ee
	so that any power of this matrix just yields the identity matrix. 
	\par
	Taking this into account and taking the sum of the first three heat kernel coefficients we get
	\bea \label{WWW}W&&=- \int d^3 x \, \sqrt{|\bg|^{(3)}} \, \dfrac{1}{(4\pi)^{3/2}} \mbox{tr}\sum_{k=-\infty}^{\infty}\Big[ \left( \dfrac{2 k \pi}{L}\right)^3 \Gamma \left(-\dfrac{3}{2} \right)a_0(\Delta)^A_A + \left( \dfrac{2 k \pi}{L}\right) \Gamma\left(-\dfrac{1}{2} \right) a_1(\Delta)^A_A \nonumber \\
	&&+ \Gamma \left(\dfrac{1}{2} \right) \left(\dfrac{L}{2 k \pi }\right)a_2(\Delta)^A_A \Big]\eea
Finally, using the gravitational equation of motion,
	\be
	\bR_{\m\n} =\dfrac{1}{2} \bg_{\m\n} + F_\Lambda (\varphi)\bg_{\m\n}  \Lambda  ,
	\ee
	the on-shell effective action reads
	\bea
	W &&=  \int d^3 x \, \sqrt{|\bg|^{(3)}} \, \left[-\frac{4 \pi ^2}{45 L^3}  -\dfrac{\Lambda}{L} \left( \frac{7}{9}  F_\Lambda(\varphi)+\frac{ F''_\Lambda(\varphi)}{12 \kappa ^2}\right)+\dfrac{L \gamma_E \Lambda^2}{\pi^2} \left( -\frac{F''_\Lambda(\varphi)^2}{16 \kappa ^4} \right. \right.\nonumber \\
	&&\left. \left.+\frac{F_\Lambda(\varphi) F''_\Lambda(\varphi)}{12 \kappa ^2}+\frac{ F'_\Lambda(\varphi)^2}{2\kappa ^2}+\frac{7}{5}  F_\Lambda(\varphi)^2\right) \right].
	\eea
	\\
	%\textcolor{red}{Jesus}
	%Finally, the effective action
	%\bea &&W=-\int d^3 x \, \sqrt{-g^{(3)}} \Bigg\{\frac{4\pi^2}{45L^3}+\frac{1}{12L}\left(\frac{28}{3}\Lambda F_{\Lambda}[\varphi]+\frac{\Lambda}{\k^2}F^{''}_{\Lambda}[\varphi]\right)+\nonumber\\
	%&&+\frac{L}{8\pi^2}\g\Big[-\frac{56}{5}\Lambda F^{2}_{\Lambda}[\varphi]-\frac{4\Lambda^2}{\k^2}\left(F^{'}_{\Lambda}[\varphi]\right)^2-\frac{2\Lambda^2}{3\k^2}F_{\Lambda}[\varphi]F^{''}_{\Lambda}[\varphi]+\frac{\Lambda^2}{2\k^4}\left(F^{''}_{\Lambda}[\varphi]\right)^2\Big]\Bigg\}\eea
	%\textcolor{red}{End Jesus}\\
	Let us comment now on the result we obtain. First of all, we focus on the leading term, which is four times the energy of a scalar field, as we could already anticipate from the counting of the degrees of freedom. But this result cannot be correct in the unimodular limit, as there's no arbitrary function that prevents the coupling of this volume term to gravity. This is due to the singular limit mentioned at the beginning of the computation. When \eqref{redef} vanishes, there is no kinetic term for the new field $\varphi$ and that leads to a non-invertible internal metric $C_{AB}$. The volume term is special because it is only dependent on the trace of the identity given by the product of the internal metric with its inverse, so this clearly fails in the unimodular limit because of the singular character of this matrix. 
	\par
	Second, we see that the subleading terms, depend on the cosmological constant and the arbitrary function in front of the original term in the action. Taking the unimodular limit, this function has to be able to cancel the square root of the determinant of the metric so that the cosmological constant does not couple to gravity in the unimodular case, as it is well-known. Nevertheless, as the unimodular limit turns out to be singular (there is no way of going to the Einstein frame), we cannot trust these results in that limit either. However, it is fortunate that at this point we can rely on an independent calculation of the vacuum energy in Unimodular Gravity by two different groups \cite{AP,Percacci:2017fsy}. Both groups show that in that case, the vacuum energy does not couple to the gravitational field, that is, it does not weigh in the same sense as all other forms of energy.
	%\textbf{PARA COMPROBAR \eqref{WWW} EN UNIMODULAR NECESITAMOS $F_\Lambda(\Omega)\equiv \Omega^n f_\Lambda (f^{-1}\left(\Omega^{n-2}\right))$ CON $\Omega(\varphi)$, PERO ESTO ULTIMO NO ES POSIBLE}
	%%%%%%%%%%%%%%%%%%%%%%%%%%%%%%%%%%%%%%%%%%%%%%%%%%%%%%%%%%%%%%%%%%%%%%%%%%%%%%%%%%%%%%%%%%%%%%%%%%%%%%%%%%%%%%%
	%%%%%%%%%%%%%%% %%%%%%%%%%%%%%%%%%%%%%%%%%%%%%%%%%%%%%%%%%%%%%%%%%%%%%%%%%%%%%%%%%%%%%%%%%%%%%%%%%%%%%%%%%%%%%%
	%%%%%%%%%%%%%%%%%%%%%%%%%%%%%%%%%%%%%%%%%%%%%%%%%%%%%%%%%%%%%%%%%%%%%%%%%%%%%%%%%%%%%%%%%%%%%%%%%%%%%%%%%%%%%%%
	%%%%%%%%%%%%%%%%%%%%%%%%%%%%%%%%%%%%%%%%%%%%%%%%%%%%%%%%%%%%%%%%%%%%%%%%%%%%%%%%%%%%%%%%%%%%%%%%%%%%%%%%%%%%%%%
	%%%%%%%%%%%%%%%%%%%%%%%%%%%%%%%%%%%%%%%%%%%%%%%%%%%%%%%%%%%%%%%%%%%%%%%%%%%%%%%%%%%%%%%%%%%%%%%%%%%%%%%%%%%%%%%
	%%%%%%%%%%%%%%%%%%%%%%%%%%%%%%%%%%%%%%%%%%%%%%%%%%%%%%%%%%%%%%%%%%%%%%%%%%%%%%%%%%%%%%%%%%%%%%%%%%%%%%%%%%%%%%%
	
	%%%%%%%%%%%%%%%%%%%%%%%%%%%%%%%%%%%%%%%%%%%%%%%%%%%%%%%%%%%%%%%%%%%%%%
	
	%%%%%%%%%%%%%%%%%%%%%%%%%%%%%%%%%%%%%%%%%%%%%%%%%%%%%%%%%%%%%%%%%%%%%%%%%%%%%%%%%%%%%%%%%%%%%%%%%%%%%%%%%%%%%%%%%%%%%%%%%%%%
	\section{Conclusions}
	%%%%%%%%%%%%%%%%%%%%%%%%%%%%%%%%%%%%%%%%%%%%%%%%%%%%%%%%%%%%%%%%%%%%%%%%%%%%%%%%%%%%%%%
	In this paper we have  discussed the quantum field vacuum energy in several  contexts. In the background field formalism
	that we use all along, vacuum energy appears as the field-independent term of the effective potential, that is, a cosmological constant. This is true no matter whether the gravitational field is considered as a non-dynamical background, or else as a quantized dynamical entity. In that sense the weight of the vacuum energy is guaranteed {\em ab initio} to be the same  as any other form of energy and no ambiguity should arise.
	\par
	We have studied spacetime manifolds of the type $\mathbb{T}_3\times \mathbb{R}$ (where the real line represents time), which are particularly interesting from the physical point of view. The general case corresponding to manifolds  of the form $\mathbb{F}_3\times \mathbb{R}$, $\mathbb{F}_3$ being flat, were completely classified by Wolf in his famous book \cite{Wolf}. 
	%and reviewed in Appendix \ref{W}.
	For the sake of brevity, we have only derived a general formula for the effective potential density of $\mathbb{T}_3\times \mathbb{R}$ manifolds, although we conjecture that our calculation could be easily extended to the other flat manifolds in Wolf's list. 
	We find a quite simple (albeit non-local) relationship between physics at radius $R$ and physics at radius $l_s^2/R$. This relationship, which ultimately stems from Poisson's summation formula and the magic of Riemann's theta functions, is somewhat similar to the one appearing in string theory. The difference is that the free energy and its dual are not proportional, but rather related through an integral transform.
	\par
	We have also studied quantum gravity corrections to the  vacuum energy and find an unambiguous energy momentum tensor for the vacuum energy. This tells us how vacuum energy {\em weighs}, in agreement with the equivalence principle, as we argued earlier on.
	It is also remarkable that the contribution of gravitons to the vacuum energy is twice the one stemming from scalars.
	This was already known in flat spacetime but we have showed that it remains true for quite general backgrounds.
	\par
	Finally, we have extended our calculation to transverse gravity, invariant under transverse diffeomorphisms only (those are the ones such that its generating vector field  is transverse, that is, $\pd_\m\xi^\m=0$.) Unfortunately our techniques fail in the most interesting case, which is the case of Unimodular Gravity. General arguments however guarantee that vacuum energy does {\em not} weigh in this case. In fact this is not exactly true, owing to self consistency imposed by Bianchi identities, but at any rate the weigh should remain independent of $\mathcal{E}_0$. 
	\par
	This is a physical prediction, which could  be verified in a laboratory. This allows Unimodular Gravity to be disproved. We are aware of the difficulties of such an experiment, but hopefully precision measurements would be carried out in the future years. One should never underestimate the ingenuity of our experimental colleagues. 
	\section{Acknowledgements}
	%%%%%%%%%%%%%%%%%%%%%%%%%%%%%%%%%%%%%%%%%%%%%%%%%%%%%%%%%%%%%%%%%%%%%%%%%%%%%%%%%%%%%%%
	This work has been partially supported by the Spanish Research Agency (Agencia Estatal de Investigacion) through the PID2019-108892RB-I00/AEI/ 10.13039/501100011033 grant as well as the IFT Centro de Excelencia Severo Ochoa SEV-2016-0597 one, and the European Union's Horizon 2020 research and innovation programme under the Marie Sklodowska-Curie grants agreement No 674896 and No 690575. RSG is supported by the Spanish FPU Grant No FPU16/01595.

	\newpage
	\appendix

	\section{Some details of the computations}\label{S}
	%%%%%%%%%%%%%%%%%%%%%%%%%%%%%%%%%%%%%%%%%%%%%%%%%%%%%%%%%%%%%%%%%%%%%%%%%%%%%%%%%%%%%%%%%%%%%%%%%%%%%%%%%%%%%%%%
	\subsection{The effect of dynamical gravity on the vacuum energy}
	For the first computation, we take the following action
	\be S=\int d^n x\sqrt{|g|}\Big[-\frac{1}{2\kappa^2}R+\frac{1}{2}\partial_\m\phi\partial^\m\phi-\frac{m^2}{2}\phi^2-\frac{\l}{4!}\phi^4\Big],\ee
	together with the classical background expansion
	\bea g_{\m\n}&&=\bg_{\m\n}+\kappa h_{\m\n}\nonumber\\
	\phi&&=\bp+\phi.\eea
	The equations of motion for this action then read 
	\bea\label{EM}&&\bR_{\m\n}-\frac{1}{2}\bg_{\m\n}\bR+\frac{\kappa^2}{2}\bg_{\m\n}\bn^\r\bp\bn_\r\bp-\kappa^2\bn_\m\bp\bn_\n\bp+\kappa^2\bg_{\m\n}\left(-\frac{m^2}{2}\bp^2-\frac{\l}{4!}\bp^4\right)=0\nonumber\\
	&&-\bar{\Box}\bp-m^2\bp-\frac{\l}{6}\bp^3=0.
	\eea
	We use a generalized De Donder gauge given by
	\be
	S_{\text{\tiny{GF}}}=\int\,d^nx\,\,\sqrt{|\bg|}\,\frac{1}{4}\,\bg_{\m\n}\chi^\m\chi^\n
	\ee
	with 
	\be
	\chi^\n=\bn_\m h^{\m\n}-\frac{1}{2}\bn^\n h-2\kappa\phi\bn^\n\bp
	\ee	
	The quadratic piece of the action, after gauge fixing, takes the form
	\be
	S_{2+gf} = \frac{1}{2}\int \, \sqrt{|\bg|} \, d^4x \, \Phi^A \Delta_{AB} \Phi^B 
	\ee
	where
	\be\label{oper}
	\Delta_{AB}=-g_{AB}\bar{\Box}+Y_{AB}
	\ee
	and
	\be
	\psi^A\equiv\begin{pmatrix}
		h^{\a\b}\\\phi
	\end{pmatrix}
	\ee
	The internal metric takes the form
	\be
	g_{AB}=\begin{pmatrix}
		C_{\a\b\m\n}&0\\0&1 \end{pmatrix}
	\ee
	with
	\bea
	C_{\m\n\r\s}&=&\frac{1}{8}(\bg_{\m\r}\bg_{\n\s}+\bg_{\m\s}\bg_{\n\r}-\bg_{\m\n}\bg_{\r\s})\nonumber\\
	C^{\m\n\r\s}&=&2(\bg^{\m\r}\bg^{\n\s}+\bg^{\m\s}\bg^{\n\r}-\frac{2}{n-2}\bg^{\m\n}\bg^{\r\s})\nonumber\\
	\eea
	The components of $Y_{AB}$ are also detailed below
	\bea
	Y^{hh}_{AB}&=&
	\frac{1}{8}(\bg_{\a\m}\bg_{\b\n}+\bg_{\a\n}\bg_{\b\m}-\bg_{\a\b}\bg_{\m\n})\bR+\frac{1}{4}\left(\bg_{\a\b}\bR_{\m\n}+\bg_{\m\n}\bR_{\a\b}\right)-\nonumber\\&&-\frac{1}{8}\left(\bg_{\a\m}\bR_{\b\n}+\bg_{\a\n}\bR_{\b\m}+\bg_{\b\m}\bR_{\a\n}+\bg_{\b\n}\bR_{\a\m}\right)-\frac{1}{4}\left(\bR_{\m\a\n\b}+\bR_{\n\a\m\b}\right)+\nonumber\\
	&&+\frac{\kappa^2}{4}\left(\bg_{\a\m}\bn_\b\bp\bn_\n\bp+\bg_{\a\n}\bn_\b\bp\bn_\m\bp+\bg_{\b\m}\bn_\a\bp\bn_\n\bp+\bg_{\b\n}\bn_\a\bp\bn_\m\bp\right)-\nonumber\\
	&&-\frac{\kappa^2}{4}\left(\bg_{\a\b}\bn_\m\bp\bn_\n\bp+\bg_{\m\n}\bn_\a\bp\bn_\b\bp\right)-\nonumber\\
	&&-\frac{\kappa^2}{4}(\bg_{\a\m}\bg_{\b\n}+\bg_{\a\n}\bg_{\b\m}-\bg_{\a\b}\bg_{\m\n})\left(\frac{1}{2}\bn_\r\bp\bn^\r\bp-\frac{m^2}{2}\bp^2-\frac{\l}{4!}\bp^4\right)\nonumber\\
	Y^{h\phi}_{AB}&=&Y^{\phi h}_{AB}=2\kappa\left(\frac{1}{2}\bn_\a\bn_\b\bp-\frac{1}{4}\bg_{\a\b}\bar{\Box}\bp-\bg_{\a\b}\frac{m^2}{4}\bp-\frac{\l}{4!}\bg_{\a\b}\bp^3\right)\nonumber\\
	Y^{\phi\phi}_{AB}&=&-m^2-\frac{\l}{2}\bp^2+2\kappa^2\bn_\r\bp\bn^\r\bp
	\eea
	The contribution coming from the ghost loops is also needed. The ghost Lagrangian is obtained performing a variation on the gauge fixing term 
	\be
	\d\chi_\n=\frac{1}{\kappa}\left(\bar{\Box}\bg_{\m\n}+\bR_{\m\n}-2\kappa^2\bn_\m\bp\bn_\n\bp\right)\xi^\m,
	\ee
	plus terms that give operators cubic in fluctuations and therefore are irrelevant at one loop. The ghost Lagrangian then reads
	\be
	S_{gh}=\frac{1}{2}\,\,\int\,d^nx\,\,\sqrt{|\bg|}\,\frac{1}{2}\,V^*_\m\left(-\bar{\Box}\bg^{\m\n}-\bR^{\m\n}+2\kappa^2\bn^\m\bp\bn^\n\bp\right)V_\n
	\ee
	With these, we can compute the traces of the different total heat kernel coefficients, where we also include the ghost contribution (the extra factor is coming from the fermion loop and from the complex character of the ghosts)
	\bea&&\mbox{tr} \Big[\left(a_0(\Delta) -2 a_0^{ghost}(\Delta)\right) \left(g_{AB}\right)^{3/2}\Big]=3\nonumber\\
	&&\mbox{tr} \Big[\left(a_1(\Delta)-2 a_1^{ghost}(\Delta)\right) \left(g_{AB}\right)^{1/2}\Big]=m^2+\left(5m^2\k^2+\frac{\l}{2}\right)\bp^2+\frac{5}{12}\k^2\l\bp^4-\frac{11}{2}\k^2\bn_\n\bp\bn^\n\bp\nonumber\\
	&&\mbox{tr} \Big[\left(a_2(\Delta)-2 a_2^{ghost}(\Delta)\right) \left(g_{AB}\right)^{-1/2}\Big]=\frac{m^4}{2}+\frac{1}{6}\left(-26m^4\k^2+3m^2\l\right)\bp^2+\nonumber\\
	&&+\left(-\frac{57}{20}m^4\k^4-\frac{55}{36}m^2\k^2\l+\frac{\l^2}{8}\right)\bp^4-\frac{\k^2\l}{40}\left(19m^2\k^2+5\l\right)\bp^6-\frac{19}{960}\k^4\l^2\bp^8-\nonumber\\
	&&-\k^2\left(4m^2+\frac{2}{3}\l\bp^2\right)\bp\bar{\Box}\bp-\frac{11}{6}m^2\k^2\bn_\m\bp\bn^\m\bp+\frac{1}{60}\left(171m^2\k^4-55\k^2\l\right)\bp^2\bn_\m\bp\bn^\m\bp+\nonumber\\
	&&+\frac{19}{80}\k^4\l\bp^4\bn_\m\bp\bn^\m\bp+2\k^2\bar{\Box}\bp\bar{\Box}\bp+\frac{203}{40}\k^2\bn_\m\bp\bn^\m\bp\bn_\n\bp\bn^\n\bp.\nonumber\\\eea
	%%%%%%%%%%%%%%%%%%%%%%%%%%%%%%%%%%%%%%%%%%%%%%%%%%%%%%%%%%%%%%%%%%%%%%%%%%%%%%%%%%%%%%%%%%%%%%%%%%%%%%%%%%%%%%%%%%%%%%%%%%%%%%%
	\subsection{Dynamical transverse gravity}
	%%%%%%%%%%%%%%%%%%%%%%%%%%%%%%%%%%%%%%%%%%%%%%%%%%%%%%%%%%%%%%%%%%%%%%%%%%%%%%%%%%%%%%%%%%%%%%%%%%%%%%%%%%%%%%
	For the computation regarding {\em TDiff} invariant theories, the starting point is the action given by
	\bea
	&& S=-{1\over 2 \kappa^2} \int d^n x \sqrt{|g|} \Big[R+2 \Lambda F_\Lambda(\varphi)-\k^2 g^{\m\n} \pd_\m \varphi \pd_\n\varphi-\k^2 F_{\phi}(\varphi) \, g^{\m\n}\pd_\m\phi\pd_\n\phi \Big].\nonumber\\\eea
	For this computations, we consider that the background value of the two scalar fields is constant and we expand the graviton in the usual way
	\bea g_{\m\n}&&=\bg_{\m\n}+\kappa h_{\m\n}\nonumber\\
	\phi&&=\bp+\phi\nonumber\\
	\varphi&&=\bar{\varphi}+\varphi.\eea
	Then, the quadratic piece of the action after gauge fixing (the De Donder gauge is enough here as the scalar fields have constant backgrounds) takes the form
	\be
	S_{2+gf} = \frac{1}{2}\int \, \sqrt{|\bg|} \, d^4x \, \Phi^A \Delta_{AB} \Phi^B,
	\ee
	where now the generalized field contains the extra scalar field
	\be
	\Phi^A\equiv\begin{pmatrix}
		h^{\a\b}\\\phi \\ \varphi,
	\end{pmatrix}
	\ee
	and the operator has again the symbolic form
	\bea
	\Delta_{AB}=\begin{pmatrix}
		-C_{\a\b\m\n} \bar{\Box}&0&0\\0&- F_\phi(\varphi) \bar{\Box} &0 \\ 0& 0& -\bar{\Box} \end{pmatrix}  + Y_{AB}.
	\eea
	In this case, the components of $Y_{AB}$ are
	\bea
	Y^{hh}_{AB}&=&
	\frac{1}{8}(\bg_{\a\m}\bg_{\b\n}+\bg_{\a\n}\bg_{\b\m}-\bg_{\a\b}\bg_{\m\n})\bR+\frac{1}{4}\left(\bg_{\a\b}\bR_{\m\n}+\bg_{\m\n}\bR_{\a\b}\right)-\nonumber\\&&-\frac{1}{8}\left(\bg_{\a\m}\bR_{\b\n}+\bg_{\a\n}\bR_{\b\m}+\bg_{\b\m}\bR_{\a\n}+\bg_{\b\n}\bR_{\a\m}\right)-\frac{1}{4}\left(\bR_{\m\a\n\b}+\bR_{\n\a\m\b}\right)+\nonumber\\
	&&+\frac{1}{4}(\bg_{\a\m}\bg_{\b\n}+\bg_{\a\n}\bg_{\b\m}-\bg_{\a\b}\bg_{\m\n})\Lambda F_{\Lambda}[\varphi]\nonumber\\
	Y^{h\varphi}_{AB}&=&Y^{\varphi h}_{AB}=-{1\over  2\kappa} \bg_{\a\b}\Lambda F^{'}_\Lambda(\bar{\varphi})\nonumber\\
	Y^{\varphi\varphi}_{AB}&=&-{1\over  \kappa^2}\Lambda F^{''}_\Lambda(\bar{\varphi}).
	\eea
	Finally, the trace of the heat kernel coefficients read
	\bea&&\mbox{tr} \Big[\left(a_0(\Delta) -2 a_0^{ghost}(\Delta)\right) \left(g_{AB}\right)^{3/2}\Big]=4\nonumber\\
	&&\mbox{tr} \Big[\left(a_1(\Delta) -2 a_1^{ghost}(\Delta)\right) \left(g_{AB}\right)^{1/2}\Big]=\frac{28}{3}\Lambda F_{\Lambda}[\varphi]+\frac{\Lambda}{\k^2}F^{''}_{\Lambda}[\varphi]\nonumber\\
	&&\mbox{tr} \Big[\left(a_2(\Delta) -2 a_2^{ghost}(\Delta)\right) \left(g_{AB}\right)^{-1/2}\Big]=-\frac{56}{5}\Lambda F^{2}_{\Lambda}[\varphi]-\frac{4\Lambda^2}{\k^2}F^{'}_{\Lambda}[\varphi]^2-\frac{2\Lambda^2}{3\k^2}F_{\Lambda}[\varphi]F^{''}_{\Lambda}[\varphi]+\frac{\Lambda^2}{2\k^4} F^{''}_{\Lambda}[\varphi]^2.\nonumber\\\eea
	\section{The dual r\^ole of the masses}\label{D1}
	%%%%%%%%%%%%%%%%%%%%%%%%%%%%%%%%%%%%%%%%%%%%%%%%%%%%%%%%%%%%%%%%%%%%%%%%%%%%%%%%%%%%%%%%%%%%%%%%%%%%%%%%%%%%%%%%
	
	Consider the operator ${\cal O}$ given by
	\be
	{\cal O}\equiv -\left( \Box-m^2\right),
	\ee
	with constant $m$. The heat kernel coefficients can be found in the literature so that the divergent piece (in $n=4$) of the operator reads
	\begin{eqnarray}
	\frac{1}{2}\log det\Delta&=& \frac{1}{n-4}\frac{1}{(4\pi)^2}\left[ -\frac{1}{6}\bar{R}M^2+\frac{1}{2} M^4 +\frac{1}{360}\left(5\bar{R}^2-2\bar{R}_{\m\n}^2+2\bar{R}_{\m\n\rho\sigma}^2\right) \right] 
	\end{eqnarray}
	%Looking up  \cite{Gilkey}, we check that what he calls $a_2$ has in it some mass terms, namely
	%\be
	%e_4={1\over 360(4\pi)^{n/2}}\left\{-60 R m^2+180 m^4+\ldots\right\}
	%\ee
	There is however another way of computing the same divergent piece of the determinant, namely, integrating the mass independently 
	%\be
	%K(m)=K(0) e^{-m^2\t}={1\over (4\pi\t)^{n/2}}\,\sum_p e_p(-\Box+m^2) \t^{p\over 2}
	%\ee
	%with
	%\be
	%\tr\, K(m=0)={1\over (4\pi\t)^{n/2}}\,\sum_p e_p(-\Box) \t^{p\over 2}
	%\ee
	%This yields
	%\bea
	%&&K(m)=K(0) e^{-m^2\t}={1\over (4\pi\t)^{n/2}}\,\sum_p e_p(-\Box) \t^{p\over 2}\sum_q {\left(-m^2 %\t\right)^q\over q!}=\nonumber\\
	%&&={1\over (4\pi\t)^{n/2}}\,\bigg\{e_0(-\Box)\left(1 -m^2\t+{m^4 \t^2\over 2}+\ldots\right)+e_2 (-\Box)\t\left(1-m^2\t+\ldots\right)+\nonumber\\
	%&&+e_4(-\Box)\t^2\left(1+\ldots\right)+\ldots\bigg\}
	%\eea
	\bea \frac{1}{2}\log det\Delta&&=-\frac{1}{2}\int\frac{d\t}{\t}\frac{1}{(4\pi\t)^{n/2}}e^{-m^2\t}\sum_{p=0}^{\infty}a_p\t^p=-\frac{1}{(4\pi)^{n/2}}\sum_{p=0}^{\infty}a_p \, m^{n-2p} \, \Gamma\left(p-\frac{n}{2}\right),\nonumber\\\eea
	so that the all the mass dependence is treated exactly. In $n=4-\e$ we have
	\bea \frac{1}{2}\log det\Delta&&=\frac{1}{n-4}\frac{1}{(4\pi)^2}\left[a_2(\bar{\Box})-M^2a_1(\bar{\Box})+\frac{1}{2}M^4a_0(\bar{\Box})\right]\eea
	The difference here is that $a_p(-\Box)$ is independent of $m$ and taking the values of the various heat kernel coefficients from the literature we get
	\bea
	a_0(\bar{\Box})&&=1\nonumber\\
	a_1(\bar{\Box})&&=\frac{1}{6}\bR\nonumber\\
	a_2(\bar{\Box})&&=\frac{1}{360}\left(5\bar{R}^2-2\bar{R}_{\m\n}^2+2\bar{R}_{\m\n\rho\sigma}^2\right). \eea
	We see that we obtain the same result using both methods.
	\section{Theta functions}\label{D3} 
	%%%%%%%%%%%%%%%%%%%%%%%%%%%%%%%%%%%%%%%%%%%%%%%%%%%%%%%%%%%%%%%%%%%%%%%%%%%%%%%%%%%%%%%%%%%%%%%%%%%%%%%%%%%%%%%
	Let us summarize the definitions and the principal properties of theta functions that are used in the paper (for an exhaustive exposition, see the classical text of Mumford \cite{Mumford}.
	%%%%%%%%%%%%%%%%%%%%%%%%%%%%%%%%%%%%%%%%%%%%%%%%%%%%%%%%%%%%%%%%%%%%%%%%%%%%%%%%%%%%%%%%%%%%%%%%%%%%%%%%%%%%%%%%
	\subsection{Poisson summation formula}\label{A}
	%%%%%%%%%%%%%%%%%%%%%%%%%%%%%%%%%%%%%%%%%%%%%%%%%%%%%%%%%%%%%%%%%%%%%%%%%%%%%%%%%%%%%%%%%%%%%%%%%%%%%%%%%%%%%%%%
	Many of the most interesting properties of the theta functions are simple consequence of  Poisson's  summation formula which states  that the sum over the integers of a function and of its Fourier transform is the same,
	\be
	\sum_{m\in\mathbb{Z}} f(m)=\sum_{n\in \mathbb{Z}} \widetilde f(n),
	\ee
	provided  we define the Fourier transform as
	\be
	\widetilde{f}(p)\equiv \int_{-\infty}^\infty \,dx\, e^{-2\pi i x p}\, f(x).
	\ee
	In order to prove Poisson's formula, let us define a new  function
	\be
	h(x)\equiv \sum_{q\in \mathbb{Z}} f(x+q),
	\ee
	it can be expanded in a Fourier series as
	\be
	h(x)\equiv \sum_{m\in \mathbb{Z}} c_m e^{2\pi i m x},
	\ee
	with coefficients
	\bea
	&&c_m\equiv \int_0^1  dx \,h(x)\, e^{-2\pi i m x}=\int_0^1  dx \,\sum_{q\in \mathbb{Z}} f(x+q)\, e^{-2\pi i m x}=\sum_{q\in \mathbb{Z}} \int_q^{q+1} dy\,f(y)\, e^{2\pi im(y-q)}=\nonumber\\
	&&=\int_{-\infty}^\infty dy\,f(y) e^{2\pi i m y} =\widetilde{f}(-m).
	\eea
	Now we have by definition
	\be
	\sum_{m\in\mathbb{Z}} f(m)=h(0),
	\ee
	and
	\be
	\sum_{m\in \mathbb{Z}} \widetilde{f}(-m)=\sum_{m\in \mathbb{Z}} c_m= h(0)
	\ee
	Let us now apply Poisson's firmula to the function
	\be
	f(x)=e^{\pi x^2\t}
	\ee
	whose Fourier transform reads
	\be
	\widetilde{f}(p)={1\over \sqrt{\t }} e^{-{\pi p^2\over \t}}
	\ee
	It follows that
	\be
	\sum_{x\in\mathbb{Z}}\,e^{\pi x^2\t}={1\over \sqrt{\t}} \sum_{p\in\mathbb{Z}}\,e^{-{\pi p^2\over \t}}
	\ee
	which is the basis of the modular properties of all theta functions.
	%%%%%%%%%%%%%%%%%%%%%%%%%%%%%%%%%%%%%%%%%%%%%%%%%%%%%%%%%%%%%%%%%%%%%%%%%%%%%%%%%%%%%%%%%%%%%%%%%%%%%%%%%%%%%%%%
	\subsection{Jacobi's theta function}\label{J}
	%%%%%%%%%%%%%%%%%%%%%%%%%%%%%%%%%%%%%%%%%%%%%%%%%%%%%%%%%%%%%%%%%%%%%%%%%%%%%%%%%%%%%%%%%%%%%%%%%%%%%%%%%%%%%%%%
	Jacobi's theta functions is defined as
	\be
	\vartheta(z|\t)\equiv \sum_{n\in \mathbb{Z}}\, e^{\pi i n^2 \t+ 2\pi i n z},
	\ee
	and obeys the differential equation given by
	\be
	{\pd \over \pd \t}\,\vartheta(z|\t)={i\over 4\pi}\,{\pd^2\over \pd z^2}\,\vartheta(z|\t).
	\ee
	This is nothing but the heat equation with proper time
	\be
	\t_{proper}={i\over 4\pi}\,\t.
	\ee
	Moreover, taking the small proper time limit we obtain
	\be
	\lim_{\t\rightarrow 0} \vartheta(z|\t)=\sum_{n\in \mathbb{Z}}\, e^{2\pi i n z}=\sum_{p\in \mathbb{Z}}\,\d(z-p).
	\ee
	\par
	A very important property of this function is the modular property. Consider $(a,b,c,d)\in \mathbb{Z}$ and such that $ad-bc=1$. Then
	\be
	\vartheta\left(\left.{z\over z \t + d}\right|{a\t+b\over c\t+d}\right)=\zeta \,(c\t+d)^{1/2}\,e^{\pi i z^2{c\over c\t+d}}\,\vartheta\left(\left.z\right|\t\right).
	\ee
	This is quite simple to prove for $\theta(0|i\t)$ by using  Poisson`s summation formula, presented in the previous section, (\ref{A}). As a particular case we have
	\be\label{tt}
	\vartheta\left(0\bigg|-{1\over \t}\right)=\t^{1/2}\,\vartheta\left(0|\t\right).
	\ee
	%Poisson's formula  (confer \cite{Koblitz}, p.72) reads
	%\be
	%\sum_{m\in\mathbb{Z}} \hat{f}(2\pi m)=\sum_{m\in\mathbb{Z}} \, f(m)
	%\ee
	%where we have defined Fourier's transform as
	%\be
	%\hat{f}(p)\equiv \int_{-\infty}^\infty \,e^{-2\pi i p x }\, f(x)\, dx
	%\ee 
	%%%%%%%%%%%%%%%%%%%%%%%%%%%%%%%%%%%%%%%%%%%%%%%%%%%%%%%%%%%%%%%%%%%%%%%%%%%%%%%%%%%%%%%%%%%%%%%%%%%%%%%%%%%%%%%%
	\subsection{Riemann theta function}\label{R}
	%%%%%%%%%%%%%%%%%%%%%%%%%%%%%%%%%%%%%%%%%%%%%%%%%%%%%%%%%%%%%%%%%%%%%%%%%%%%%%%%%%%%%%%%%%%%%%%%%%%%%%%%%%%%%%%%%
	The Riemann theta function is a generalization of the Jacobi theta function. Taking 
	\be\mathbb{H}_n=\left\{F\in M(n,\mathbb{C}) \,\middle|\, F=F^\mathsf{T} \,\,\operatorname{Im} F >0 \right\},\ee
	to be the set of symmetric square matrix whose imaginary part is positive definite, and given $\Omega \in \mathbb{H}_n$ the Riemann theta function is defined as
	
	\be \Theta \left(z\middle|\Omega\right)=\sum_{m\in \mathbb{Z}^g}
	\exp\left(2\pi i \left(\frac{1}{2} m^\mathsf{T} \Omega m +m^\mathsf{T} z \right)\right)
	\ee
	here, $z \in\mathbb{C}^g$ is an g-dimensional complex vector, and the superscript $T$ denotes the transpose. By construction, the Riemann theta function is periodic in $(z-z^\prime)$
	\be
	\Theta\left(z-z^\prime\middle|\Omega\right)= \Theta\left(z-z^\prime+m\middle| \Omega\right)
	\ee
	for arbitrary $m\in \mathbb{Z}^g$.
	
	The modular property reads \cite{Mumford}
	\be \Theta\left(\left[[C\Omega\t+D]^{-1}\right]^{
		\mathrm{T}}\cdot\mathbf{z}\middle|[A\Omega\t+B][
	C\Omega\t+D]^{-1}\right)=t_\g\sqrt{\det[C\Omega\t+D]}e^{\pi i%
		\mathbf{z}\cdot\left[[[C\Omega\t+D]^{-1}C\right]\cdot\mathbf{z}}\Theta\left(\mathbf{z}\middle|\Omega\t
	\right),\ee
	% \be\vartheta\left(\,\Big[\left(C\Omega+ D\right)^{-1}\Big]^T.\vec{x}| \left(A\Omega+B\right)\left(C\Omega+D\right)^{-1}\right)=t_\g \sqrt{\det\,\left(C\Omega+D\right)}\,e^{\pi i \,t \vec{x}.\left(C \Omega+D\right)^{-1} C \vec{x}}\,\vartheta\left(\vec{x},\Omega\right)
	%\ee
	where $(t_\g)^8 =1$ and $\g\equiv \bpm A&B\\C&D\epm \in Sp(4,\mathbb{Z})$. 
	%%%%%%%%%%%%%%%%%%%%%%%%%%%%%%%%%%%%%%%%%%%%%%%%%%%%%%%%%%%%%%%%%%%%%%%%%%%%%%%%%%%%%%%%%%%%%%%%%%%%%%%%%%%%%%%%
	\subsection{Dimensional reduction and oxidation}\label{D2}
	%%%%%%%%%%%%%%%%%%%%%%%%%%%%%%%%%%%%%%%%%%%%%%%%%%%%%%%%%%%%%%%%%%%%%%%%%%%%%%%%%%%%%%%%%%%%%%%%%%%%%%%%%%%%%%%%%
	Consider  a scalar field in a gravitational background as the one considered previously
	\be
	S=\int d^4x\sqrt{|g|}\, \phi(x,y)\Box_4 \phi(x,y).
	\ee
	Working on a manifold of the form $M_4 \equiv M_{3}\times S^1$, we can expand in the field in harmonics
	\be
	\phi(x,y)=\frac{1}{\sqrt{L}}\phi_n(x)\,e^{i n 2 \pi y /L},
	\ee
	%\be
	%\int_0^{2\pi L} dy\, e^{i m 2 \pi y /L}\,e^{i n 2 \pi y /L}=2\pi L\, \d(m+n)
	%\ee
	so that the quadratic part of the action reads (after the integration of the compact dimension)
	\be
	S=\sum_n \int d^3x\sqrt{|g|^{(3)}}\, \phi_n(x)\Big[\Box_3+ \left(\frac{2\pi n}{L}\right)^2\Big]\phi_n(x)
	\ee
	
	Our aim is to show that when $L\rightarrow 0$ (reduction) the theory reduces to a three-dimensional one, and that when $L\rightarrow\infty$ the theory cannot be told apart from the ordinary four-dimensional one (oxidation). %The short time expansion of the heat kernel is given by
	%\be
	%K(x,\xp;\t)={1\over (4\pi\t)^{n/2}}e^{-{\s(x,\xp)\over 2\t}}\,\sum_{n=0}^\infty a_n \t^{n/2}
	%\ee
	%On the other hand, the relationship
	%\be
	%a_k=\sum_{p+q=k} a_p(M_3)a_q(S^1)
	%\ee
	%taking into account that for $S^1$ (of unit radius)
	%\be
	%\tr\, K(\t)=\sum e^{-\t n^2}\quad\Longrightarrow \bigg\{ a_0(S^1)=(4\pi)^{- n/2}.\quad a_{k\neq 0} (S^1)=0\bigg\}
	%\ee
	%does not yield much information.
	The heat kernel we are interested in can be factorized as
	\bea
	\tr K_{M_3\times S^1}(\t)&&=\tr\,K_{M_3}\left(\t\right)\vartheta \left(0\middle|{i \t\over \pi L^2}\right)=\tr\,K_{M_3}\left(\t\right)\left({i\pi L^2\over \t }\right)^{1/2}\,\vartheta\left(0\middle|i\pi{ \ L^2\over \t}\right),\nonumber\\
	\eea
	where we have used the property \eqref{tt} in the last equality.

	%For a general manifold (including $M_3\times S^1$)  we know that
	%\bea
	%a_0&&={1\over (4\pi)^{n/2}}\nonumber\\
	%a_1&&={1\over 6}{1\over (4\pi)^{n/2}}\bigg\{6 E+ R\bigg\}\nonumber\\
	%a_2&&={1\over 360}{1\over (4\pi)^{n/2}}\bigg\{60\Box E+60 E R+ 180 E^2+12\Box R+5 R^2+2 R_{\m\n\r\s}^2\bigg\}
	%\eea
	The problem is how to recover four-dimensional results out of three-dimensional ones. Reduction is easy, because
	\be
	\lim_{L\rightarrow 0} \vartheta\left(0\bigg|{i\t\over \pi L^2}\right)=1
	\ee
	\bea
	&&\tr K_{M_3\times S^1}(\t)=\tr K_{M_3}
	\eea
	Oxidation is also clear, just because we also have
	\be
	\lim_{L\rightarrow\infty} \vartheta\left(0\left.\right|i\pi L^2\right)=1,
	\ee
	and then,
	\bea
	&&\tr K_{M_3\times S^1}(\t)=\tr\,K_{M_3}\left(\t\right)\left({i\pi L^2\over \t }\right)^{1/2}.
	\eea
	It would be interesting to discover the physical interpretation of the factor $\left({i\pi L^2\over \t }\right)^{1/2}$.
	
	%%%%%%%%%%%%%%%%%%%%%%%%%%%%%%%%%%%%%%%%%%%%%%%%%%%%%%%%%%%%%%%%%%%%%%%%%%%%%%%%%%%%%%%%%%%%%%%%%%%%%%%%%%%%%%%
	\section{Regularization of $\zeta(1)$} \label{E}
	%%%%%%%%%%%%%%%%%%%%%%%%%%%%%%%%%%%%%%%%%%%%%%%%%%%%%%%%%%%%%%%%%%%%%%%%%%%%%%%%%%%%%%%%%%%%%%%%%%%%%%%%%%%%%
	Throughout the text, we make use of the zeta function regularization in various computations. In this appendix we go through some of the details used for the $\zeta(1)$ case. We take as the starting point the sum given by
	\be
	\sum_{n=1}^{n=\infty }{1\over n}\equiv \lim_{\e\rightarrow 0}\sum_{n=1}^{n=\infty }{1\over n}e^{-\e n},
	\ee
	and define
	\be
	S(\e)\equiv \sum_{n=1}^{n=\infty }{1\over n}\,e^{-\e n}.
	\ee
	Taking a first derivative of this function we obtain
	\bea
	&&{d S(\e)\over d \e}=-\sum_{n=1}^{n=\infty }\,e^{-\e n}={1\over 1- e^\e},
	\eea
	so that we can further write 
	\be
	S(\e)=\e-\log\,(e^\e-1)+C.
	\ee
	Taking the limit when $\e\rightarrow 0$, we finally get
	\be
	S(\e)\sim \log\,|\e|+C
	\ee
	Let us note that we have implemented the boundary condition
	\be
	\lim_{\e\rightarrow\infty} S(\e)=0.
	\ee
	%We shall determine the constant $C$ in a moment.
	%\par
	%Also
	%\be
	%\sum_{n=1}^\infty 1\equiv \lim_{\e\downarrow 0}{d S(\e)\over d \e}=-\lim_{\e\downarrow 0}{1\over e^\e-1}={1\over \e}-{1\over 2}
	%\ee
	%\be
	%\sum_{n=1}^\infty n\equiv \lim_{\e\downarrow 0}{d^2 S(\e)\over d \e^2}
	%\ee
	%\be
	%\sum_{n=1}^\infty n^2\equiv \lim_{\e\downarrow 0}{d^3 S(\e)\over d \e^3}=-{2\over \e^3}
	%\ee
	%In all cases the finite part coincides with the value given by Riemann's  $\zeta$-function. The last evaluation is a manifestation of the fact that
	%\be
	%\zeta(-2k)=0
	%\ee
	%for $k\in \mathbb{N}$.
	\par
	We can now determine the constant $C$ taking
	\be
	S(0)= \sum_{n=1}^\infty {1\over n}=\zeta(1)=\infty
	\ee
	which does not seem to help. Nevertheless, near $s=1$ on the real axis
	\be
	\zeta(s)={1\over s-1}+\g_E-\g_1 (s-1)+O(s-1)^2
	\ee
	where $\g_E=0.5772$ is Euler's Gamma constant and $\g_1=-0.0728$ is Stieljes' constant. This  is also true going along  the imaginary axis
	\be
	\zeta(1+i\e)={1\over i \e}+\g_E-i \g_1 \e+O(\e^2),
	\ee
	so that we can take the finite value of $\zeta(1)=\gamma_E.$
	\par
	\newpage
	%%%%%%%%%%%%%%%%%%%%%%%%%%%%%%%%%%%%%%%%%%%%%%%%%%%%%%%%%%%%%%%%%%%%%%%%%%%%%%%%%%%%%%%%%%%%%%%%%%%%%%%%%%%%%%%%%
	
\end{document}